\documentclass[epsfig,12pt]{article}
\usepackage{makeidx}
\usepackage{amsmath}
\usepackage{amsfonts}
\usepackage{amssymb}
\usepackage{graphicx}
\usepackage{accents}

\input epsf.sty
\newtheorem{theorem}{Theorem}
\newtheorem{acknowledgement}[theorem]{Acknowledgement}

\makeatletter \@addtoreset{equation}{section}

\input{tcilatex}
\begin{document}

\title{\rightline{\mbox{\small
{Lab/UFR-HEP0805/GNPHE/0805}}} \textbf{On} \textbf{Black Hole Effective
Potential in }\\
\textbf{6D/7D }$\mathcal{N}\mathbf{=2}$\textbf{\ Supergravity}}
\author{ El Hassan Saidi{\small \thanks{%
h-saidi@fsr.ac.ma}} \\
{\small 1. Lab/UFR- Physique des Hautes Energies, Facult\'{e} des Sciences,
Rabat, Morocco,}\\
{\small 2. Groupement National de Physique des Hautes Energies, Si\`{e}ge
focal: FS, Rabat}\\
{\small 3. Coll\`{e}ge SPC, Acad\'{e}mie Hassan II des Sciences et
Techniques, Rabat, Morocco} }
\maketitle

\begin{abstract}
Using the harmonic superspace method and the duality between real and
complex representations of hypermultiplets, we compute the explicit scalar
field expression of the quaternionic metric $G_{mn}\left( \varphi \right) $
of the moduli space $\frac{SO\left( 4,k\right) \times SO\left( 1,1\right) }{%
SO\left( 4\right) \times SO\left( k\right) }$ of $6D$ $\mathcal{N}\mathbf{=2}
$ supergravity with generic $k$ Maxwell supermultiplets. The obtained metric
includes the particular case $k=20$ associated with 10D type IIA superstring
on K3. Uplifting to $7D$ $\mathcal{N}\mathbf{=2}$ supergravity is described
and aspects of $6D/7D$\ black attractor effective potentials are studied.\ 
\newline
\textbf{Key words}: Type IIA superstring on K3, Quaternionic geometry, 6D
black attractors, Uplifting to 7D.
\end{abstract}


\section{Introduction}

The dynamics of \emph{10D} type II superstrings compactified on the real
four dimensional K3 manifold at Planck scale is described by \emph{6D} $%
\mathcal{N}=2$ supergravity \textrm{\cite{W}-\cite{HT}}.\ One distinguishes
two 6D models, A and B, depending on whether one started from \emph{10D}
type IIA or \emph{10D} type IIB superstrings\textrm{.} These models are
respectively given by the usual \emph{6D} non chiral $\mathcal{N}=\left(
1,1\right) $ and \emph{6D} chiral $\mathcal{N}=\left( 2,0\right) $
supersymmetric models and have different moduli spaces $\boldsymbol{M}_{%
\text{\textsc{IIA/K3}}}=\frac{SO\left( 4,20\right) \times SO\left(
1,1\right) }{SO\left( 4\right) \times SO\left( 20\right) }$ and $\boldsymbol{%
M}_{\text{\textsc{IIB/K3}}}=\frac{SO\left( 5,21\right) }{SO\left( 5\right)
\times SO\left( 21\right) }$ with real dimensions $\dim \boldsymbol{M}_{%
\text{\textsc{IIA/K3}}}=80+1$ and $\dim \boldsymbol{M}_{\text{\textsc{IIA/K3}%
}}=105$ respectively.

In this paper, we focus our attention on the study the interacting dynamics
of the \emph{81} scalar fields $\varphi ^{m}=\varphi ^{m}\left( x\right) $
of the non chiral \emph{6D} $\mathcal{N}=2$ supergravity. Our interest in
this issue has been motivated by looking for the \emph{explicit} field
expression of the \emph{6D} black hole effective potential $\mathcal{V}%
_{eff}^{6D,N=2}\left( \varphi \right) $ which depends on the scalar fields
self couplings.\newline
Generally, the interacting dynamics of the $\varphi ^{m}$'s is described by
the following typical non linear sigma model field action,%
\begin{equation}
\mathcal{S}_{b}\left[ \varphi \right] =\int d^{6}x\sqrt{-\det g}\left(
\sum\limits_{\mu ,\nu =0}^{5}g^{\mu \nu }\left[ \sum\limits_{m,n=1}^{\dim 
\boldsymbol{M}}\frac{\partial \varphi ^{m}}{\partial x^{\mu }}\frac{\partial
\varphi ^{m}}{\partial x^{\nu }}\hat{G}_{mn}\right] \right) ,  \label{1}
\end{equation}%
where $\boldsymbol{M}$ stands for the scalar manifold $\boldsymbol{M}_{\text{%
\textsc{IIA/K3}}}$. The above field action $\mathcal{S}_{b}\left[ \varphi %
\right] $ appears in \emph{6D} $\mathcal{N}=2$ supergravity as the scalar
field part of the action $\mathcal{S}_{\text{6D sugra}}\left[ \varphi ,...%
\right] $ describing the dynamics of all the degrees of freedom of the
theory.\newline
The main objective of this study is to use the link between quaternionic
geometry and \emph{6D} supersymmetry to determine the explicit scalar field
expression of the self couplings matrix,%
\begin{equation}
\begin{tabular}{llll}
$\hat{G}_{mn}$ & $=$ & $\hat{G}_{mn}\left( \varphi \right) $ & ,%
\end{tabular}
\label{gg}
\end{equation}%
of the field action $\mathcal{S}_{b}\left[ \varphi \right] $. We also take
this opportunity to give results regarding the \emph{family} of the
quaternionic scalar manifolds $\frac{SO\left( 4,k\right) }{SO\left( 4\right)
\times SO\left( k\right) }$ with $k\geq 1$, concerning generic \emph{6D} $%
\mathcal{N}=2$ supergravity models as well as the \emph{cousin family} $%
\frac{SO\left( 3,k\right) }{SO\left( 3\right) \times SO\left( k\right) }$
dealing with uplifting to \emph{7D} space time\textrm{\footnote{%
Following \cite{la} and using the decomposition $so\left( 4,20\right)
=so\left( 3,19\right) +so\left( 1,1\right) +\left( 3,19\right) ^{+}+\left(
3,19\right) ^{-}$, the quaternionic manifold $\frac{SO\left( 4,20\right) }{%
SO\left( 4\right) \times SO\left( 20\right) }$\ can be realized as a
fibration over the submanifold $\frac{SO\left( 3,19\right) }{SO\left(
3\right) \times SO\left( 19\right) }$. }}.\newline
Notice that besides the fact that the computation of $\hat{G}_{mn}$ is by
itself an interesting question, the knowledge of its explicit relation in
terms of the field variables $\left\{ \varphi ^{m}\right\} $ is particularly
important for the study of black attractors in \emph{6D} and \emph{7D} space
time dimensions \textrm{\cite{BDSS}-\cite{HH}; }see also\textrm{\ \cite{F3}-%
\cite{F31}}. The role of the metric $\hat{G}_{mn}$ in this matter should be
compared with the role played by the Kahler metric%
\begin{equation}
g_{i\overline{j}}\sim \frac{\partial ^{2}\mathcal{K}\left( z,\overline{z}%
\right) }{\partial z^{i}\partial \overline{z}^{\overline{j}}}\qquad ,\qquad 
\mathcal{K}=\mathcal{K}\left( z,\overline{z}\right) \text{ ,}  \label{gk}
\end{equation}%
in the study of the BPS and non BPS attractors in \emph{4D} $\mathcal{N}=2$
supergravity \textrm{\cite{BA1}-\cite{BA4}}; and, up on imposing some
constraint eqs, in the uplifting to $\mathcal{N}=2$ supergravity in \emph{5D}
space time \textrm{\cite{5D1}}-\textrm{\cite{5D4}}. \newline
To fix the ideas, recall that in \emph{10D} type IIA superstring on K3,
which is dual to the \emph{10D} heterotic superstring on the real 4- torus $%
\mathbb{T}^{4}$, the scalar manifold $\boldsymbol{M}$ is given by the
following particular non compact real space 
\begin{equation*}
\begin{tabular}{llllllll}
$\boldsymbol{M}$ & $=$ & $\boldsymbol{Q}_{80}\times SO\left( 1,1\right) $ & ,
& $\dim \boldsymbol{M}$ & $=$ & $81$ & ,%
\end{tabular}%
\end{equation*}%
where the factor $\boldsymbol{Q}_{80}$ is the real \emph{eighty} dimensional
quaternionic space, 
\begin{equation}
\begin{tabular}{llllllll}
$\boldsymbol{Q}_{80}$ & $\boldsymbol{=}$ & $\frac{SO\left( 4,20\right) }{%
SO\left( 4\right) \times SO\left( 20\right) }$ & , & $\dim \boldsymbol{Q}%
_{80}$ & $=$ & $4\times 20$ & ,%
\end{tabular}
\label{ms}
\end{equation}%
with the isotropy symmetry $SO\left( 4\right) \times SO\left( 20\right) $.%
\newline
The real one dimensional $SO\left( 1,1\right) $ factor is parameterized by
the dilaton $\sigma $; which generally appears in the analysis as a
multiplicative factor $e^{-m\sigma }$ with some number $m$. If freezing the
dynamics of $\sigma $ by imposing the constraint relation, 
\begin{equation}
\begin{tabular}{llll}
$d\sigma $ & $=$ & $0$ & ,%
\end{tabular}
\label{si}
\end{equation}%
then the dynamics of the scalars of the \emph{6D} supergravity theory
reduces to that of the \emph{eighty} scalars $\left\{ \phi ^{m}\left(
x\right) \right\} $ parameterizing $\boldsymbol{Q}_{80}$. In this case, the $%
\hat{G}_{mn}\left( \sigma ,\phi \right) $ coupling matrix of eq(\ref{1})
reduces as well to the restricted field coupling $G_{mn}=G_{mn}\left( \phi
\right) $ which is nothing but the field metric of $\boldsymbol{Q}_{80}$,
that is: 
\begin{equation*}
\begin{tabular}{llll}
$dl^{2}$ & $=$ & $\tsum\limits_{m,n=1}^{\dim \boldsymbol{Q}}G_{mn}d\phi
^{m}d\phi ^{m}$ & .%
\end{tabular}%
\end{equation*}%
For later use, it is useful to rewrite the above relation in a more
convenient way by taking advantage of the isometries of the scalar manifold $%
\boldsymbol{Q}_{80}$ \textrm{\cite{S1,BDSS}}. Then, we have, 
\begin{equation}
\begin{tabular}{llll}
$dl^{2}$ & $=$ & $\tsum\limits_{a,b=1}^{4}\tsum%
\limits_{I,J=1}^{20}G_{ab}^{IJ}d\phi _{I}^{a}d\phi _{J}^{b}$ & ,%
\end{tabular}
\label{dl}
\end{equation}%
where the real \emph{80} field variables $\phi _{I}^{a}$ are in the
bi-fundamental of $SO\left( 4\right) \times SO\left( 20\right) $ isotropy
symmetry and where $G_{ab}^{IJ}=G_{ab}^{IJ}\left( \phi \right) $ is the
metric of $\boldsymbol{Q}_{80}$ that we want to determine explicitly; but
now expressed in the local coordinate frame $\left\{ \phi _{I}^{a}\right\} $.%
\newline
As noticed above, the knowledge of the \emph{explicit} field expression of
the coupling $G_{ab}^{IJ}$ is crucial in the study of the BPS and non BPS\
black attractors in \emph{6D} supergravity. There, the black hole (and up on
using electric/magnetic duality the black membrane) effective potential $%
\mathcal{V}_{eff}^{6D,N=2}\left( \phi \right) $ has the form \textrm{\cite%
{S1,F2},}%
\begin{equation}
\begin{tabular}{llll}
$\mathcal{V}_{eff}^{6D,N=2}\left( \phi \right) $ & $=$ & $%
\sum\limits_{a,b=1}^{4}\mathcal{K}^{ab}\left(
Z_{a}Z_{b}+\sum\limits_{I,J=1}^{20}G_{ab}^{IJ}Z_{I}Z_{J}\right) \geq 0$ & ,%
\end{tabular}
\label{ve}
\end{equation}%
where $Z_{a}=Z_{a}\left( \phi \right) $ and $Z_{I}=Z_{I}\left( \phi \right) $
are respectively the so called geometric and matter central charges of the 
\emph{6D} $\mathcal{N}=2$ supersymmetric algebra and where the factor $%
G_{ab}^{IJ}$ is as in eq(\ref{dl}). \newline
The above effective potential (\ref{ve}) should be compared with the well
known potential of the black hole in 4D $\mathcal{N}=2$ supergravity \textrm{%
\cite{BFKM}},%
\begin{equation}
\begin{tabular}{llll}
$\mathcal{V}_{eff}^{4D,N=2}\left( z,\overline{z}\right) $ & $=$ & $e^{%
\mathcal{K}}\left( \left\vert \mathcal{Z}\right\vert
^{2}+\sum\limits_{i,j=1}^{m}\mathrm{g}^{i\overline{j}}\mathcal{Z}_{i}%
\overline{\mathcal{Z}}\overline{_{j}}\right) \geq 0$ & ,%
\end{tabular}%
\end{equation}%
where $\mathcal{K}$ and $\mathrm{g}_{i\overline{j}}$ are as in eq(\ref{gk}).%
\newline
Moreover, by focusing on eq(\ref{ve}), we see that the knowledge of $%
\mathcal{V}_{eff}^{6D,N=2}$ requires, in addition to $Z_{a},$ $Z_{I}$ and
the matrix potential $\mathcal{K}^{ab}$, the metric $G_{ab}^{IJ}$ in terms
of the fields $\phi ^{aI}$. With the explicit field expression of eq(\ref{ve}%
) at hand, one can write down the explicit field expression of the attractor
eqs of \emph{6D} black attractor. These eqs are given by the criticality
condition, 
\begin{equation*}
\frac{\partial \mathcal{V}_{eff}\left( \phi \right) }{\partial \phi ^{aI}}=0%
\text{ .}
\end{equation*}%
The explicit fields coupling $G_{ab}^{IJ}$ is also useful for elaborating
generalizations beyond the \emph{6D} supergravity limit of \emph{10D} type
IIA superstring on K3 ; in particular for the two following issues:\newline
(\textbf{a}) \emph{reducing or} \emph{adding extra quaternionic dimensions}: 
\newline
(\textbf{i}) the dimension reduction of $Q_{80}$\ down to $Q_{4k}$ with $%
k<20 $ can be done by looking at particular realizations of the compact K3
manifold. This restriction corresponds to setting part of the moduli to zero 
\begin{equation*}
\begin{tabular}{llll}
$\left\{ \phi ^{m}\right\} _{1\leq m\leq k-1}$ & , & $\left\{ \phi
^{m}\right\} _{k\leq m\leq 20}\rightarrow 0$ & ,%
\end{tabular}%
\end{equation*}%
leading then to K3's with singularities \textrm{\cite{BBS1}-\cite{BBDS}}. 
\newline
(\textbf{ii}) the extension of $Q_{80}$ up to higher dimensional $Q_{4k}$
with $k>20$ can be done by going beyond the \emph{6D} supergravity models
that follow from \emph{10D} type IIA superstring on K3. In this case, we
have, in addition to the eighty field variables, extra quaternionic moduli. 
\newline
These two situations correspond to \emph{6D} supergravity theories with
generic moduli spaces 
\begin{equation*}
\begin{tabular}{lllll}
$Q_{4k}$ & $=$ & $\frac{SO\left( 4,k\right) }{SO\left( 4\right) \times
SO\left( k\right) }$ & $k\geq 1$ & ,%
\end{tabular}%
\end{equation*}%
and make the results to be derived throughout this study more general.%
\newline
Moreover, seen that the scalar manifolds of the 6D$\ \mathcal{N}=2$
supergravity models are real \emph{4n} dimensional manifolds, one may use
other Lie group representations such as the real \emph{4n} dimensional
symplectic coset $\frac{SP\left( 2,2n\right) }{SP\left( 2\right) \times
SP\left( 2n\right) }$. However, it turns out to be more interesting to,
instead of real field coordinates $\left\{ \phi \right\} $, use the $2n$
local complex coordinates $\left\{ \mathrm{f}^{1A},\overline{\mathrm{f}}%
_{1A};\mathrm{f}^{2A},\overline{\mathrm{f}}_{2A}\right\} _{A=1,...,n}$ by
thinking about the real scalar manifolds $Q_{4n}$ as follows,%
\begin{equation}
\begin{tabular}{lllllll}
$\boldsymbol{H}_{2n}$ & $\boldsymbol{=}$ & $\frac{SU\left( 2,n\right) }{%
SU\left( 2\right) \times U\left( n\right) }$ & , & $\dim _{R}\boldsymbol{H}%
_{2n}$ & $=$ & $2\times \left( 2n\right) $%
\end{tabular}%
.  \label{nh}
\end{equation}%
The quaternionic manifolds $\boldsymbol{H}_{2n}$, which are contained%
\footnote{%
Below, we shall think about $Q_{4n}$, with real field coordinates $\phi
_{I}^{a}$, and $H_{2n}$, with complex fields $\mathrm{f}^{iA}$, as roughly
referring to the same scalar manifold of the \emph{6D} $\mathcal{N}=2$
supergravity. The two field coordinates are related by duality
transformations (\ref{chh}). The metric of $Q_{4n}$ is denoted as G$%
_{ab}^{IJ}$ and often used to refer to the metric of $H_{2n}$ which has
three component blocks $\left( h_{kB}^{iA},g^{iAkB},\overline{g}%
_{iAkB}\right) $ as in eq(\ref{ss}).} in $Q_{4n}$, define as well an
infinite family of real $4n$- (complex $2n$- ) dimensional spaces. \newline
Notice that for the leading $n=1$ term of this series, the manifold $%
\boldsymbol{H}_{2}$ has an $SU\left( 2\right) \times U\left( 1\right) $
isotropy symmetry. This isometry can be remarkably interpreted as the same
isometry that we have in the real 4 dimensional Taub-Newman-Unti-Tamburino
(Taub-NUT) metric \textrm{\cite{Ta,N,GM,IS}}. The latter has been
extensively studied in literature from several views; it will be revisited
in section 4 from the view of 6D black hole perspective. \newline
(\textbf{b}) \emph{Uplifting to 7D}\newline
By borrowing the idea of 4D/5D correspondence of $\mathrm{\cite{5D2}}$, the
results obtained for the case 6D black attractors could a priori be used to
derive their 7D counterpart. The uplifting to 7D is too particularly
interesting in dealing with the effective potential $\mathcal{V}%
_{eff}^{7D,N=2}$ of black attractors in \emph{7D} $\mathcal{N}=2$
supergravity and the classification of \emph{7D} BPS and non BPS solutions
along the line of \textrm{\cite{5D2}}. The first step in this way is the
determination of the metric $\left( G_{\alpha \beta }^{uv}\right) _{1\leq
\alpha ,\beta \leq 3}^{1\leq u,v\leq 19}$ of the generic moduli space, 
\begin{equation}
\begin{tabular}{llllll}
$\boldsymbol{M}_{n}^{7D}$ & $=$ & $\frac{SO\left( 3,n\right) \times SO\left(
1,1\right) }{SO\left( 3\right) \times SO\left( n\right) }$ & , & $n\geq 1$ & 
.%
\end{tabular}%
\end{equation}%
Recall that for $n=19$, the \emph{7D} $\mathcal{N}=2$ supergravity is a
limit of \emph{11D} M- theory on K3. This theory can be also recovered by
uplifting the \emph{6D} $\mathcal{N}=2$ supergravity to \emph{7D}. \newline
The moduli space $\boldsymbol{M}_{19}^{7D}$ can be obtained from eq(\ref{ms}%
) by switching off the \emph{22} fluxes of the \emph{NS-NS} B- field over
the 2-cycles of K3; together with a constraint on the volume of K3. Using
this property, one can \`{a} priori determine the explicit field metric $%
\left( G_{\alpha \beta }^{uv}\right) _{1\leq \alpha ,\beta \leq 3}^{1\leq
u,v\leq 19}$ of the generic spaces $\boldsymbol{M}_{n}^{7D}$ by imposing
appropriate constraint eqs on the metric $G_{ab}^{IJ}$ of the moduli space $%
\boldsymbol{M}_{n}^{6D}$ of the \emph{6D} supergravity. A comment regarding
this issue will be made in the discussion section.

The organization of this paper is as follows: In section 2, we review
briefly the fields content of \emph{10D} type IIA on K3 and recall useful
aspects of harmonic superspace (\emph{HSS}) method. In section 3, we focus
on the scalars of the \emph{6D} theory. We develop a duality transformation
mapping real coordinates $\phi _{k}^{ia}$ complex ones f$^{iA}$ and use it
to study the abelian gauge invariance of the Maxwell-matter sector as well
as their self couplings by using results on 6D supersymmetry in harmonic
superspace. In section 4, we compute the explicit expression of the metric
components $h_{kB}^{iA},$ $g^{iAkB}$ and $\overline{g}_{iAkB}$ ( $%
G_{ab}^{IJ} $ for short) of the scalar manifolds $\boldsymbol{H}_{2n}$ (\ref%
{nh}). In section 5, we give the conclusion and make two discussions; one on
the explicit field expression of the 6D black hole potential and the other
regarding the uplifting to 7D. In the appendices 6 and 7, we give
respectively some useful tools on harmonic superspace and on the geometric
approach of $\widehat{\mathrm{F}}_{4}$ supergravity.

\section{Fields in 10D type IIA on K3}

In this section, we review briefly some useful results on \emph{10D} type
IIA superstring on K3. We begin by recalling the two sectors of the spectrum
of the \emph{10D} type IIA superstring: \newline
(\textbf{1}) the perturbative sector containing the following \emph{10D} IIA
supergravity massless fields, 
\begin{equation}
\begin{tabular}{llllllll}
\emph{NS-NS} {\small fields} & $:\quad $ & $g_{MN}$ & , & $B_{MN}$ & , & $%
\phi _{dil}$ & , \\ 
\emph{RR} {\small fields \ \ \ \ \ } & $:\quad $ & $\mathcal{A}_{M}$ & , & $%
\mathcal{C}_{MNK}$ &  &  & ,%
\end{tabular}
\label{tds}
\end{equation}%
where the index $M.=0,\cdots ,9$ captures the 10- vectors of $SO\left(
1,9\right) $. \newline
Along with these bosonic fields, which carry a total number of $128$ on
shell degrees of freedom, we also have two \emph{10D}- gravitinos and two 
\emph{10D}- gauginos. \newline
(\textbf{2}) the non perturbative sector containing D- branes carrying RR
charges. They are collected in the following table together with the
associated gauge invariant field strengths,

\begin{equation}
\begin{tabular}{llllllllll}
Type IIA D- branes & $:\quad $ & $D0$ & , & $D2$ & , & $D4$ & , & $D6$ & ,
\\ 
Field strenghts \ \ \ \ \ \  & $:\quad $ & $\mathcal{F}_{2}$ & , & $\mathcal{%
F}_{4}$ & , & $^{\ast }\mathcal{F}_{4}$ & , & $^{\ast }\mathcal{F}_{2}$ & ,%
\end{tabular}%
\end{equation}%
where $\ast $ stands for the Hodge dual.

\subsection{Compactification on K3}

Under the compactification of \emph{10D} type IIA superstring on K3, the $%
SO\left( 1,9\right) $ space-time symmetry breaks down to the subgroup $%
SO\left( 1,5\right) \times SU_{R}\left( 2\right) $; which is contained in $%
SO\left( 1,5\right) \times SO\left( 4\right) $. Moreover, the initial the 
\emph{32} conserved supersymmetric charges get reduced down to \emph{16}. 
\newline
The degrees of freedom of \emph{10D} type IIA superstring on K3 describe, at
the gravity level, a non chiral \emph{6D} $\mathcal{N}=2$ supergravity
theory and appear in two irreducible $\mathcal{N}=2$ supersymmetric
representations namely the gravity supermultiplet and Maxwell ones. Below,
we describe these supermultiplets.\newline

(\textbf{1}) \emph{6D }$\mathcal{N}=2$\emph{\ gravity supermultiplet} \
\qquad\ \newline
The bosonic fields of the six dimensional $\mathcal{N}=2$ gravity
supermultiplet contains $32$ on shell degrees of freedom distributed as 
\begin{equation}
\begin{tabular}{llllll}
$g_{\mu \nu }$ & , & $\mathcal{B}_{\mu \nu }$ & , & $\sigma $ & , \\ 
$\mathcal{A}_{\mu }^{\left( ij\right) }$ & , & $\mathcal{C}_{\mu \nu \rho }$
&  &  & ,%
\end{tabular}
\label{gr}
\end{equation}%
where $\mu ,\nu =0,...,5$ stands for the space-time indices and $i,j=1,2$
are the isospin 1/2 indices of the $SU_{R}\left( 2\right) $ symmetry. 
\newline
The $g_{\mu \nu }$ is the space time metric and $\mathcal{B}_{\mu \nu }$ the 
\emph{6D} antisymmetric 2-form.\newline
The gauge fields $\mathcal{A}_{\mu }^{\left( ij\right) }$ and $\mathcal{C}%
_{\mu \nu \rho }$ can be thought of as \emph{four} gravi-photons.\newline
Besides the real 3-form $\mathcal{H}_{3}=d\mathcal{B}$, we also have the
following the gauge invariant field strengths, 
\begin{equation}
\begin{tabular}{llll}
$\mathcal{F}_{2}^{\left( ij\right) }$ & $=$ & $d\mathcal{A}_{1}^{\left(
ij\right) }$ & , \\ 
$\mathcal{F}_{4}^{0}$ & $=$ & $d\mathcal{C}_{3}$ & ,%
\end{tabular}
\label{fc}
\end{equation}%
and their duals%
\begin{equation}
\begin{tabular}{llll}
$\mathcal{F}_{4}^{\left( ij\right) }$ & $=$ & $\text{ }^{\ast }\mathcal{F}%
_{2}^{\left( ij\right) }$ & , \\ 
$\mathcal{F}_{2}^{0}$ & $=$ & $\text{ }^{\ast }\mathcal{F}_{4}^{0}$ & .%
\end{tabular}
\label{fcc}
\end{equation}%
Notice the two following features:\newline
(\textbf{a}) the \emph{6D} fields $\mathcal{A}_{\mu }^{\left( ij\right) }$, $%
\mathcal{F}_{2}^{\left( ij\right) }$ and $^{\ast }\mathcal{F}_{2}^{\left(
ij\right) }$ are $SU_{R}\left( 2\right) $ isotriplets while $\mathcal{C}_{3}$%
, $\mathcal{F}_{4}^{0}$ and $\mathcal{F}_{2}^{0}$ are isosinglets. \newline
(\textbf{b}) the 6D gauge field $\mathcal{C}_{3}$ is dual to a \emph{6D}
Maxwell field $\mathcal{A}_{\mu }^{0}$. \newline
So eqs(\ref{fc}-\ref{fcc}) can be also exhibited in terms of quartets as
follows%
\begin{equation}
\begin{tabular}{llll}
$\mathcal{A}_{\mu }^{ij}$ & $=$ & $\mathcal{A}_{\mu }^{0}\mathcal{%
\varepsilon }^{ij}+\mathcal{A}_{\mu }^{\left( ij\right) }$ & ,%
\end{tabular}
\label{red}
\end{equation}%
and 
\begin{equation}
\begin{tabular}{llll}
$\mathcal{F}_{2}^{ij}$ & $=$ & $d\mathcal{A}^{ij}$ & , \\ 
$\mathcal{F}_{4}^{ij}$ & $=$ & $\text{ }^{\ast }\mathcal{F}_{2}^{ij}$ & ,%
\end{tabular}%
\end{equation}%
with $\mathcal{\varepsilon }^{ij}=-\mathcal{\varepsilon }^{ji}$ and $%
\mathcal{\varepsilon }^{12}=1$. Notice that the fields $\mathcal{F}_{2}^{ij}$
and $\mathcal{F}_{4}^{ij}$ can be also decomposed as in eq(\ref{red}). 
\newline

(\textbf{2}) \emph{6D }$\mathcal{N}=2$\emph{\ Maxwell supermultiplets}$:$%
\emph{\ }$\left( \mathcal{V}_{6D}^{N=2}\right) _{I}$\newline
The Maxwell-matter sector of the \emph{6D} $\mathcal{N}=2$ supergravity
theory embedded in \emph{10D} type IIA superstring on K3 involves \emph{%
twenty} \emph{6D} Maxwell supermultiplets%
\begin{equation}
\begin{tabular}{llll}
$\mathcal{V}_{6D,N=2}^{I}$ & $:\qquad $ & $I=1,...,20$ & .%
\end{tabular}%
\end{equation}%
Each supermultiplet $\mathcal{V}_{6D,N=2}$ has $\left( 8+8\right) $ on shell
degrees of freedom. The \emph{eight} bosonic degrees of freedom are captured
by a \emph{6D} gauge field $\mathcal{A}_{\mu }$ and \emph{four} real scalars 
$\phi ^{ij}$. The \emph{eight} fermionic degrees of freedom are captured by
two spinors $\lambda ^{1}$ and $\lambda ^{2}$:%
\begin{equation}
\begin{tabular}{llllll}
$\left[ \mathcal{V}_{6D,N=2}\right] _{\text{Bose}}$ & $=$ & $\mathcal{A}%
_{\mu }$ & $\oplus $ & $\phi ^{ij}$ & , \\ 
$\left[ \mathcal{V}_{6D,N=2}\right] _{\text{Fermi}}$ & $=$ & $\mathcal{%
\lambda }_{\hat{\alpha}}^{{\small 1}}$ & $\oplus $ & $\mathcal{\lambda }_{%
\hat{\alpha}}^{{\small 2}}$ & .%
\end{tabular}
\label{af}
\end{equation}%
These fields transform in different representations of the $SU_{R}\left(
2\right) $ symmetry. The gauge field $\mathcal{A}_{\mu }$ is a isosinglet,
the two gauginos $\mathcal{\lambda }_{\hat{\alpha}}^{i}$ form an isodoublet
and the four scalars form as a reducible quartet; that is 
\begin{equation}
\begin{tabular}{llllll}
$4$ & $=$ & $1$ & $+$ & $3$ & .%
\end{tabular}%
\end{equation}%
In the \emph{6D} field theory set up, the four scalars are described by the
sum of a singlet $\phi ^{0}$ and a triplet $\phi ^{\left( ij\right) }$ as
shown below,%
\begin{equation}
\begin{tabular}{llllll}
$\phi ^{ij}$ & $=$ & $\phi ^{0}\varepsilon ^{ij}$ & $+$ & $\phi ^{\left(
ij\right) }$ & .%
\end{tabular}%
\end{equation}%
We will need this property later on when we consider the geometric
interpretation of the $\phi ^{ij}$'s as periods of a quaternionic 2-form $%
\boldsymbol{J}^{ij}$ to be introduced at appropriate time.\newline
Notice that the Maxwell supermultiplet $\mathcal{V}_{6D,N=2}$ contains
scalar fields that allow to make a formal correspondence with the Coulomb
branch in the $\mathcal{N}=2$ supergravity theory in $4D$ space time. 
\newline
Notice also that generally, the \emph{twenty} Maxwell supermultiplets $%
\mathcal{V}_{6D,N=2}^{I}$; in particular their Bosonic sector 
\begin{equation}
\begin{tabular}{llllll}
$\left[ \mathcal{V}_{6D,N=2}^{I}\right] _{Bose}$ & $=$ & $\mathcal{A}_{\mu
}^{I}$ & $\oplus $ & $\phi ^{ijI}$ & ,%
\end{tabular}%
\end{equation}%
the field components have quantum numbers with respect to the $SO\left(
4\right) \times SO\left( 20\right) $ isotropy symmetry of the moduli space (%
\ref{ms}). We have%
\begin{equation}
\begin{tabular}{llll}
$\mathcal{A}_{\mu }^{I}$ & $\simeq $ & $\left( \underline{1},\underline{2}%
0\right) $ & , \\ 
$\phi ^{ijI}$ & $\simeq $ & $\left( \underline{4},\underline{20}\right) $ & ,%
\end{tabular}
\label{fi}
\end{equation}%
where the $SO\left( 4\right) $ isotropy is thought of as $SU_{R}\left(
2\right) \times SU_{R}\left( 2\right) $. We also have for the gauge field
strengths%
\begin{equation}
\begin{tabular}{lllll}
$\mathcal{F}_{2}^{I}$ & $=d\mathcal{A}^{I}$ & $\quad \simeq \quad $ & $%
\left( \underline{1},\underline{2}0\right) $ & , \\ 
$\mathcal{F}_{4}^{I}$ & $=\text{ }^{\ast }\mathcal{F}_{2}^{I}$ & $\quad
\simeq \quad $ & $\left( \underline{1},\underline{2}0\right) $ & .%
\end{tabular}%
\end{equation}

\subsection{$\mathcal{N}=1$ formalism in \emph{6D}}

To study the geometry of the scalar manifold (\ref{ms}), it is enough to
focus the attention on the scalar fields $\left\{ \phi ^{ijI}\right\} $.
This can be nicely done by using \emph{6D} $\mathcal{N}=1$ supersymmetric
representations by splitting the \emph{6D} $\mathcal{N}=2$ gauge multiplet $%
\mathcal{V}_{6D}^{N=2}$ as the sum of two $6D$ $\mathcal{N}=1$ multiplets;
namely a vector multiplet $V_{6D}^{\mathcal{N}=1}$ and a hypermultiplet $%
\mathcal{H}_{6D}^{\mathcal{N}=1}$,%
\begin{equation}
\begin{tabular}{llllll}
$\mathcal{V}_{6D}^{N=2}$ & $=$ & $\mathcal{V}_{6D}^{N=1}$ & $\oplus $ & $%
\mathcal{H}_{6D}^{N=1}$ & ,%
\end{tabular}
\label{spl}
\end{equation}%
with%
\begin{equation}
\begin{tabular}{llllll}
$V_{6D}^{\mathcal{N}=1}$ & $=$ & $\left( \mathcal{A}_{\mu },\mathcal{\lambda 
}_{\hat{\alpha}}\right) $ & $\equiv $ & $\left( 1,\frac{1}{2}\right) _{6D}$
& , \\ 
$\mathcal{H}_{6D}^{\mathcal{N}=1}$ & $=$ & $\left( \phi ^{ij},\mathcal{\psi }%
_{\hat{\alpha}}\right) $ & $\equiv $ & $\left( 0^{4},\frac{1}{2}\right)
_{6D} $ & ,%
\end{tabular}
\label{vh}
\end{equation}%
where $1,$ $\frac{1}{2}$ and $0$ stand for the space time spin of the
component fields and the powers for their numbers. \newline
Notice the three following features:\newline
First, the decomposition (\ref{spl}) is a general property of $\mathcal{N}=2$
supersymmetry in any space-time dimension. Irreducible supermultiplets $%
\left( R_{\mathcal{N}=2}\right) $ can be usually split into pairs of $%
\mathcal{N}=1$ irreducible representations as given below, 
\begin{equation}
\begin{tabular}{llllll}
$R_{\mathcal{N}=2}$ & $=$ & $R_{\mathcal{N}=1}$ & $\oplus $ & $R_{\mathcal{N}%
=1}^{\prime }$ & .%
\end{tabular}%
\end{equation}%
Second, the vector supermultiplet $\mathcal{V}_{6D}^{\mathcal{N}=1}$ has a
gauge field but no scalars; while the hypermultiplet $\mathcal{H}_{6D}^{%
\mathcal{N}=1}$ has no vector field but four scalars capturing the
quaternionic structure of the Coulomb branch of \emph{6D} $\mathcal{N}=2$
supersymmetry. \newline
Finally, it is interesting to note that there is a remarkable parallel
between the reductions of the irreducible vector representations of $%
\mathcal{N}=2$ supersymmetry in \emph{6D} and \emph{4D} space times, 
\begin{equation}
\begin{tabular}{llllll}
$\mathcal{N}=2$ & $\rightarrow $ & $\mathcal{N}=1$ & $\oplus $ & $\mathcal{N}%
^{\prime }=1$ & .%
\end{tabular}%
\end{equation}%
Concerning the underlying geometries of the associated Coulomb branches, we
have

\begin{equation}
\begin{tabular}{|l|l|l|l|}
\hline
{\small vector multiplet} &  & {\small matter multiplet} & {\small scalar
manifold} \\ \hline
$4D$ $\mathcal{N}=2$ & $\rightarrow $ & $\mathcal{N}=1$ {\small chiral matter%
} & {\small Kahler} \\ \hline
$6D$ $\mathcal{N}=2$ & $\rightarrow $ & $\mathcal{N}=1$ {\small hyper matter}
& {\small quaternionic} \\ \hline
\end{tabular}%
\end{equation}

\ \ \newline
In superspace formulation of $\mathcal{N}=1$ supersymmetry in \emph{4D}, the
scalar multiplet $\Phi _{4D}^{N=1}$ is described by a chiral superfield $%
\Phi =\Phi \left( x,\theta \right) $ with dynamics described by the
superspace Lagrangian density,%
\begin{equation}
\begin{tabular}{lllll}
$\mathcal{L}_{4D}^{N=1}$ & $=$ & $\ \ \ \ \ \ \int d^{4}\theta \text{ }%
\mathcal{K}\left( \Phi ,\bar{\Phi}\right) $ &  &  \\ 
&  & $+$ $\ \ \int d^{2}\theta \text{ }W\left( \Phi \right) $ $\ \ \ +$ & $%
\int d^{2}\bar{\theta}\text{ }\bar{W}\left( \bar{\Phi}\right) $ & ,%
\end{tabular}
\label{kp}
\end{equation}%
where $\mathcal{K}\left( \Phi ,\bar{\Phi}\right) $ is the Kahler potential
and $W\left( \Phi \right) $ the chiral superpotential .\newline
In the harmonic superspace(\emph{HSS}) formulation of $\mathcal{N}=1$
supersymmetry in \emph{6D}, the hypermultiplet is described by the
superfield $\Phi ^{+}=\Phi ^{+}\left( x,\theta ^{+},u^{\pm }\right) $ with
dynamics governed a by the \emph{HSS}\ Lagrangian density%
\begin{equation}
\begin{tabular}{llll}
$\mathcal{L}_{6D}^{+4}$ & $=$ & $\int d^{4}\theta ^{+}du\left[ \tilde{\Phi}%
^{+}D^{++}\Phi ^{+}+\mathcal{L}_{int}^{+4}\left( \Phi ^{+},\tilde{\Phi}%
^{+}\right) \right] $ & ,%
\end{tabular}
\label{l}
\end{equation}%
where 
\begin{equation}
\begin{tabular}{llll}
$D^{++}$ & $=$ & $\partial ^{++}-2\theta ^{+\alpha }\theta ^{+\beta
}\partial _{\left[ \alpha \beta \right] }$ & , \\ 
$\partial ^{++}$ & $=$ & $u^{+i}\frac{\partial }{\partial u^{-i}}$ & ,%
\end{tabular}
\label{dpp}
\end{equation}%
is the \emph{HSS} covariant derivatives and $\partial _{\left[ \alpha \beta %
\right] }=\frac{\partial }{\partial x^{\left[ \alpha \beta \right] }}\sim 
\frac{\partial }{\partial x^{\mu }}$. \newline
The superfield $\mathcal{L}_{int}^{+4}$ is the \emph{HSS} potential which
can be thought of as the quaternionic superpotential that specify the
geometry of the quaternionic scalar manifold. \newline
Below, we will refer to $\mathcal{L}_{int}^{+4}$ as the quaternionic
potential.\newline
Notice that the charges $q$ carried by the HSS superfunction $F^{q}$ are the
charges of the $U_{C}\left( 1\right) $ Cartan subgroup of the $SU_{R}\left(
2\right) $ symmetry. We have%
\begin{equation*}
\left[ D^{0},F^{q}\right] =qF^{q}
\end{equation*}%
where $D^{0}$ is the generators of $U_{C}\left( 1\right) $. The operator $%
D^{0}$ together with the covariant derivatives $D^{++}$ of eq(\ref{dpp}) and
its adjoint $D^{--}$ are the generators of the $SU_{R}\left( 2\right) $
symmetry satisfying the usual commutation relation%
\begin{equation*}
\begin{tabular}{llll}
$\left[ D^{0},D^{++}\right] $ & $=$ & $+2D^{++}$ & , \\ 
$\left[ D^{0},D^{--}\right] $ & $=$ & $-2D^{--}$ & , \\ 
$\left[ D^{++},D^{--}\right] $ & $=$ & $+D^{0}$ & .%
\end{tabular}%
\end{equation*}%
More details can be found in the original works on \emph{HSS }\textrm{\cite%
{1,2}}; some useful relations are collected in the appendix of this paper.

\section{$U^{20}\left( 1\right) $ invariance and quaternionic potential}

\qquad In this section, we consider the two main points:\newline
(\textbf{1}) the field theoretic implementation of the $U^{20}\left(
1\right) $ gauge invariance of the matter sector of the \emph{10D} type IIA
superstring on K3.\newline
(\textbf{2}) the derivation of the explicit expression of the quaternionic
potential $\mathcal{L}_{int}^{+4}$ (\ref{1}) of the scalar manifold $%
\boldsymbol{H}_{2n}$ (\ref{nh}). \newline
To achieve these goals, we shall use:\newline
(\textbf{a}) tools on the real second homology/cohomology of K3; \newline
(\textbf{b}) known results on the \emph{HHS} method for hyperKahler metrics
building.

This section is organized in three subsections: \newline
In the first subsection, we first study the two following things:\newline
(\textbf{i}) develop two \emph{dual }descriptions of 6D\ hypermultiplets; 
\newline
the first description involves real scalars $\left\{ \phi ^{aI}\sim \phi
^{ijI}\right\} $ and is adapted to deal with the manifold $\frac{SO\left(
4,20\right) }{SO\left( 4\right) \times SO\left( 20\right) }$. \newline
the other realization uses complex fields $\mathrm{f}^{iA}$ and $\overline{%
\mathrm{f}}_{iA}$ concerns the complex scalar manifold $\boldsymbol{H}_{2n}$%
. \newline
(\textbf{ii}) give an heuristic derivation the quaternionic potential $%
\mathcal{L}_{int}^{+4}$ of the scalar manifold of the 10D type IIA
superstring on K3.\newline
In the two other subsections \emph{3.2} and \emph{3.3}, we give rigorous
details an the refining of the results given in subsection \emph{3.1}.

\subsection{Duality relation and quaternionic potential}

\qquad Our interest in exhibiting explicitly the $U^{20}\left( 1\right) $
symmetry is because of the central role it plays in computing the explicit
field expression of the quaternionic metric $G_{ab}^{IJ}$. Though the
existence of this symmetry is directly identified from the spectrum of the
Maxwell-matter sector of the \emph{10D} type IIA superstring on K3, 
\begin{equation}
\begin{tabular}{llllllll}
$\mathcal{A}_{\mu }^{I}$ & , & $\mathcal{\lambda }_{\alpha }^{iI}$ & , & $%
\phi ^{ijI}$ & ; & $I=1,...,20$ & ,%
\end{tabular}
\label{su}
\end{equation}%
the field theory implementation of this gauge invariance is a little bit
subtle. The point is that the scalar moduli $\phi _{k}^{iI}$ describing the
6D matter are real fields%
\begin{equation}
\begin{tabular}{llll}
$\overline{\left( \phi _{k}^{iI}\right) }$ & $=$ & $\phi _{i}^{kI}$ & ,%
\end{tabular}
\label{re}
\end{equation}%
and so neutral under $U^{20}\left( 1\right) $, 
\begin{equation}
\begin{tabular}{lllllll}
$U^{20}\left( 1\right) $ & $:$ & $\phi _{k}^{iI}$ & $\rightarrow $ & $\left(
\phi _{k}^{iI}\right) ^{\prime }=$ & $\phi _{k}^{iI}$ & .%
\end{tabular}
\label{tra}
\end{equation}%
It is then interesting to look for a dual complex description where the 
\emph{twenty} quartets of scalar fields $\phi _{k}^{iI}=\phi _{k}^{iI}\left(
x\right) $ are put into \emph{twenty} complex isodoublets 
\begin{equation}
\begin{tabular}{llllll}
$\mathrm{f}^{iA}$ & $=$ & $\mathrm{f}^{iA}\left( x\right) $ & , & $%
A=1,...,20 $ & .%
\end{tabular}
\label{g}
\end{equation}%
These complex fields allow the following phases changes%
\begin{equation}
\begin{tabular}{llll}
$\mathrm{f}^{iA\prime }$ & $=$ & $e^{i\lambda }\mathrm{f}^{iA}\left(
x\right) $ & ,%
\end{tabular}
\label{gc}
\end{equation}%
with $\lambda $ being a real diagonal matrix which can be expanded as 
\begin{equation}
\begin{tabular}{llll}
$\lambda $ & $=$ & $\tsum\limits_{I=1}^{20}\lambda _{I}T^{I}$ & .%
\end{tabular}%
\end{equation}%
In above expansion, the $T^{I}$'s are \emph{20} commuting matrices 
\begin{equation}
\begin{tabular}{llllll}
$\left\{ T^{I}\right\} _{I=1,...,20}$ & , & $T^{I}T^{J}$ & $=$ & $T^{J}T^{I}$
& ,%
\end{tabular}
\label{t}
\end{equation}%
generating the $U^{20}\left( 1\right) $ gauge invariance of the Maxwell
sector of the 6D supergravity theory.\newline

\emph{Duality relation and its superfield extension}\newline
The duality relation that maps the real fields $\phi _{i}^{kI}$ into the
complex isodoublets $\mathrm{f}^{iA}$ and $\overline{\mathrm{f}}_{iA}$ is
given by%
\begin{equation}
\begin{tabular}{llllll}
$\phi _{i}^{kI}$ & $=$ & $\overline{\mathrm{f}}_{k}T^{I}\mathrm{f}^{i}$ & $=$
& $\tsum\limits_{A,B=1}^{20}\overline{\mathrm{f}}_{kB}\left( T^{I}\right)
_{A}^{B}\mathrm{f}^{iA}$ & ,%
\end{tabular}
\label{gf}
\end{equation}%
This relation has been motivated from the relation existing between real $4$%
- vectors and the hermitian $2\times 2$ matrices; it will be described with
some details in a moment. But before coming to that, let us recall that the $%
6D$ $\mathcal{N}=2$ supergravity embedded in \emph{10D} type IIA superstring
on K3 has \emph{24} Maxwell gauge fields: \newline
- \emph{Four} of these gauge fields, denoted above as $\mathcal{A}_{\mu
}^{ij}$, belongs to the supergravity multiplet and are not directly our
target here. Nevertheless, keep in mind that the associated gauge invariant
charges are involved in the effective potential $\mathcal{V}_{eff}^{6D}$ of
the black attractor. \newline
- \emph{Twenty} other gauge fields, denoted as $\mathcal{A}_{\mu }^{I}$,
belong to the Maxwell-matter sector (\ref{su}). They transform under the
change eq(\ref{gc}) as follows,%
\begin{equation}
\begin{tabular}{lllll}
$U^{20}\left( 1\right) :$ & $\mathcal{A}_{\mu }^{I}$ & $\rightarrow $ & $%
\mathcal{A}_{\mu }^{I}+\partial \vartheta _{I}$ & ,%
\end{tabular}
\label{gi}
\end{equation}%
where the $\vartheta _{I}$'s are gauge parameters.\newline
It is these \emph{20} gauge fields and their supersymmetric partners (\ref%
{su}) that we are interested in here.\newline
Notice that the duality relation (\ref{gf}) concerning the component scalar
fields, is very suggestive. \newline
First, it can be extended to a superfield duality relation involving the
superfields 
\begin{equation}
\begin{tabular}{llll}
$H_{k}^{iI}$ & $=$ & $H_{k}^{iI}\left( x,\theta \right) $ & , \\ 
$\overline{\left( H_{k}^{iI}\right) }$ & $=$ & $H_{i}^{kI}$ & $,$%
\end{tabular}%
\end{equation}%
and 
\begin{equation}
\begin{tabular}{llll}
$\mathrm{\Phi }^{iA}$ & $=$ & $\mathrm{\Phi }^{iA}\left( x,\theta \right) $
& , \\ 
$\overline{\left( \mathrm{\Phi }^{iA}\right) }$ & $=$ & $\mathrm{\bar{\Phi}}%
_{iA}$ & ,%
\end{tabular}%
\end{equation}%
with leading $\theta $- components%
\begin{equation}
\begin{tabular}{llll}
$\phi _{i}^{kI}$ & $=$ & $\left( H_{k}^{iI}\right) _{\theta =0}$ & , \\ 
$\mathrm{f}^{iA}$ & $=$ & $\left( \mathrm{\Phi }^{iA}\right) _{\theta =0}$ & 
.%
\end{tabular}%
\end{equation}%
The superfield extension of the duality relation (\ref{gf}) reads as follows%
\begin{equation}
\begin{tabular}{llll}
$H_{k}^{iI}$ & $=$ & $\overline{\mathrm{\Phi }}_{k}T^{I}\mathrm{\Phi }^{i}$
& ,%
\end{tabular}
\label{fh}
\end{equation}%
Second, eq(\ref{fh}) has a nice description in the 6D $\mathcal{N}=1$
harmonic superspace formalism. There, the superfields $H^{ijI}$ and $\mathrm{%
\Phi }^{iA}$ are mapped to \emph{HSS} superfields 
\begin{equation}
\begin{tabular}{llll}
$H^{++I}$ & $=$ & $H^{++I}\left( x,\theta ^{+},u^{\pm }\right) $ & , \\ 
$\Phi _{A}^{+}$ & $=$ & $\Phi _{A}^{+}\left( x,\theta ^{+},u^{\pm }\right) $
& ,%
\end{tabular}%
\end{equation}%
with leading components 
\begin{equation}
\begin{tabular}{llll}
$\left( H_{I}^{++}\right) _{\theta =0}$ & $=$ & $\sum%
\limits_{i,j=1}^{2}u_{i}^{+}u_{j}^{+}\phi _{I}^{ij}$ & , \\ 
$\left( \Phi _{A}^{+}\right) _{\theta =0}$ & $=$ & $\sum%
\limits_{i=1}^{2}u_{i}^{+}\phi _{A}^{i}$ & .%
\end{tabular}%
\end{equation}%
Thus, the superfield duality relation (\ref{fh}) reads in \emph{HSS} like,%
\begin{equation}
\begin{tabular}{llll}
$H^{++I}$ & $=$ & $\mathrm{\tilde{\Phi}}^{+}T^{I}\mathrm{\Phi }^{+}$ & ,%
\end{tabular}
\label{312}
\end{equation}%
Notice in passing that the $\Phi _{A}^{+}$'s are the superfields that
describe hypermultiplets in harmonic superspace. \newline
Third, eq(\ref{312}) obey the HSS\ relation, 
\begin{equation}
\begin{tabular}{llll}
$D^{++}H^{++I}$ & $=$ & \textrm{0} & ,%
\end{tabular}
\label{313}
\end{equation}%
where the harmonic covariant derivative $D^{++}$ is as in eq(\ref{l}). These
relations have interpretation in \emph{HSS} formulation as the conservation
laws of Noether \emph{HSS} currents $\mathcal{J}^{++I}\left( x,\theta
,u^{\pm }\right) =\mathrm{\tilde{\Phi}}^{+}T^{I}\mathrm{\Phi }^{+}$.\newline

\emph{Quaternionic superpotential}\newline
Using the above tools, we can give the explicit superfield expression of the
quaternionic potential $\mathcal{L}_{int}^{+4}$ associated with the scalar
manifold $\boldsymbol{Q}_{80}$. It reads in the 6D $\mathcal{N}=1$ harmonic
superspace formalism as follows,%
\begin{equation}
\begin{tabular}{llll}
$\mathcal{L}_{n}^{+4}$ & $\simeq $ & $\frac{\lambda }{2}\tsum%
\limits_{I,J=1}^{n}\left( \tilde{\Phi}^{+}T^{I}\Phi ^{+}\right) d_{IJ}\left( 
\tilde{\Phi}^{+}T^{J}\Phi ^{+}\right) $ & ,%
\end{tabular}
\label{qp}
\end{equation}%
with $n=20$. In this relation, the real symmetric tensor $d_{IJ}=d_{JI}$ is
a coupling metric which can be interpreted in terms of intersections of
2-cycles of K3. \newline
Notice that eq(\ref{qp}) is valid for $n=20$; but also for generic integers $%
n$; in particular for $n=1$ where the above quaternionic potential reduces
to,\ 
\begin{equation}
\begin{tabular}{llll}
$\mathcal{L}_{1}^{+4}$ & $\simeq $ & $\frac{\lambda }{2}\left( \tilde{\Phi}%
^{+}\Phi ^{+}\right) ^{2}$ & .%
\end{tabular}
\label{tn}
\end{equation}%
which, according to \textrm{\cite{IS}}, is nothing but the quaternionic
potential of the real 4- dimensional Taub-NUT geometry.

\subsection{Geometric and stringy interpretations}

\qquad First notice that the real \emph{eighty} scalars $\phi _{I}^{a}=\phi
_{I}^{a}\left( x\right) $ of the 6D $\mathcal{N}=2$ supergravity are in the $%
\left( \underline{4},\underline{20}\right) $ bi-fundamental of the $SO\left(
4\right) \times SO\left( 20\right) $ isotropy symmetry of the moduli space $%
\boldsymbol{Q}_{80}$. These fields can be also written as 
\begin{equation}
\begin{tabular}{llllll}
$\phi _{I}^{a}\left( x\right) $ & $=$ & $\tsum\limits_{i,j=1}^{2}\mathcal{%
\sigma }_{ij}^{a}\phi _{I}^{ij}\left( x\right) $ & , & $I=1,...,20$ & ,%
\end{tabular}%
\end{equation}%
where $\mathcal{\sigma }^{a}$, $a=1,2,3$, are the usual $2\times 2$ Pauli
matrices and $\mathcal{\sigma }^{0}\equiv \mathcal{I}_{id}$ is the identity
matrix. Sometimes we also refer to $\mathcal{\sigma }^{0}$\ as $\mathcal{%
\sigma }^{4}$.

Viewed from the \emph{10D} type IIA superstring on K3, the \emph{80} scalar
fields $\phi _{I}^{ij}$ have two origins: \newline
- \emph{58} geometric moduli having as well an interpretation in \emph{7D} $%
\mathcal{N}=2$ supergravity.\newline
- \emph{22} stringy origin moduli; but having no analogue in \emph{7D }$%
\mathcal{N}=2$ supergravity.\newline
Let us give some details about these scalars.

(\textbf{a}) \emph{Geometric moduli}\newline
The \emph{58 }scalars\emph{\ }of the \emph{80} moduli decompose as 
\begin{equation}
\begin{tabular}{llllll}
$58$ & $=$ & $1$ & $+$ & $3\times 19$ & ,%
\end{tabular}%
\end{equation}%
and have a geometric interpretation in terms of the Kahler and complex
deformations of the metric of K3. In general, we have (\emph{1+19}) real
moduli and \emph{19} complex ones.\newline
These fields can be denoted altogether as%
\begin{equation}
\begin{tabular}{llllll}
$\phi ^{0}$ & $\oplus $ & $\phi _{I}^{\left( ij\right) }$ & , & $I=1,...,19$
& ,%
\end{tabular}
\label{19}
\end{equation}%
and belong to two kinds of representations of the $SU_{R}\left( 2\right) $
symmetry. The field variable $\phi ^{0}$ is an \emph{isosinglet}; it is
interpreted as the volume of K3. \newline
The fields $\phi _{I}^{\left( ij\right) }$ describe \emph{nineteen}
isotriplets combining the Kahler and complex deformations.

(\textbf{b}) \emph{Stringy moduli}\newline
The remaining \emph{22} field moduli decompose as 
\begin{equation}
\begin{tabular}{llllll}
$22$ & $=$ & $3$ & $+$ & $1\times 19$ & ,%
\end{tabular}
\label{20}
\end{equation}%
and have stringy interpretation in terms of the NS-NS 2-form periods. These
fields can be denoted like,%
\begin{equation}
\begin{tabular}{llllll}
$\chi ^{\left( ij\right) }$ & $\oplus $ & $\chi _{I}^{0}$ & , & $I=1,...,19$
& ,%
\end{tabular}
\label{21}
\end{equation}%
that is an isotriplet $\chi ^{\left( ij\right) }$ and \emph{nineteen}
isosinglets $\chi _{I}^{0}$.

(\textbf{c}) \emph{Comments}\newline
We give three comments.\newline
(\textbf{i}) Eqs(\ref{19}) and (\ref{20}) combine altogether into \emph{%
twenty} quartets as follows%
\begin{equation}
\begin{tabular}{llllll}
$\chi ^{ij}$ & $\oplus $ & $\phi _{I}^{ij}$ & , & $I=1,...,19$ & ,%
\end{tabular}
\label{23}
\end{equation}%
and will be read now on like, 
\begin{equation}
\begin{tabular}{llll}
$\phi _{I}^{ij}$ & , & $I=1,...,20$ & ,%
\end{tabular}%
\end{equation}%
with $\chi ^{ij}=\phi _{20}^{ij}=\phi ^{0}\varepsilon ^{ij}+\chi ^{ij}$.%
\newline
(\textbf{ii}) The fields moduli $\chi ^{0I}$ and $\phi ^{\left( ij\right) I}$
can be interpreted in terms of the periods 
\begin{equation}
\begin{tabular}{lllllll}
$\chi ^{0I}$ & $=$ & $\int_{C^{I}}\mathcal{B}_{NS}$ & , & $I$ & $=1,...,19$
& , \\ 
$\phi ^{\left( ij\right) I}$ & $=$ & $\int_{C^{I}}\boldsymbol{J}^{\left(
ij\right) }$ &  &  &  & ,%
\end{tabular}
\label{jij}
\end{equation}%
where the \emph{nineteen }$C^{I}$'s form 2-cycles sub-basis of the \emph{22}
dimensional second real homology of K3.\newline
Now, let us introduce the quaternionic 2-form%
\begin{equation}
\begin{tabular}{llll}
$\boldsymbol{J}^{ij}$ & $=$ & $\boldsymbol{J}^{\left[ ij\right] }+%
\boldsymbol{J}^{\left( ij\right) }$ & ,%
\end{tabular}
\label{22}
\end{equation}%
where $\boldsymbol{J}^{\left[ ij\right] }$ is an isosinglet 2- form and $%
\boldsymbol{J}^{\left( ij\right) }$ an isotriplet 2- form and use the $%
SU_{R}\left( 2\right) $ quantum numbers of the moduli (\ref{19}-\ref{21}) to
identify the two irreducible components of (\ref{22}). We distinguish two
representations of $\boldsymbol{J}^{ij}$ depending on the 2- cycles of K3.
To that purpose, it is interesting use the $\left( \underline{3},\underline{%
19}\right) $ signature of $H_{2}\left( K3,\mathbb{R}\right) $ to split the
real \emph{22} dimensional basis 
\begin{equation}
\begin{tabular}{llll}
$\left\{ B^{\Lambda }\right\} $ & $=$ & $\left\{ B^{1},...,B^{22}\right\} $
& ,%
\end{tabular}%
\end{equation}%
like 
\begin{equation*}
\begin{tabular}{llll}
$\left\{ B^{\Lambda }\right\} $ & $\equiv $ & $\left\{ C^{I}\right\} \oplus
\left\{ D^{a}\right\} $ & ,%
\end{tabular}%
\end{equation*}%
where, roughly, 
\begin{equation}
\begin{tabular}{llll}
$\left\{ C^{I}\right\} $ & $\sim $ & $\left\{ B^{1},...,B^{19}\right\} $ & ,
\\ 
$\left\{ D^{a}\right\} $ & $\sim $ & $\left\{ B^{20},B^{21},B^{22}\right\} $
& .%
\end{tabular}
\label{cd}
\end{equation}%
By duality 
\begin{equation}
\begin{tabular}{llll}
$\int_{B^{\Lambda }}\mathbf{\alpha }_{\Sigma }$ & $=$ & $\delta _{\Lambda
}^{\Sigma }$ & ,%
\end{tabular}%
\end{equation}%
the real 2-forms basis $\mathbf{\alpha }_{\Sigma }$ can be also split as%
\begin{equation}
\begin{tabular}{lllll}
$\left\{ \mathbf{\alpha }_{\Lambda }\right\} _{1\leq \Lambda \leq 22}$ & $%
\equiv $ & $\left\{ \mathbf{\gamma }_{I}\right\} _{1\leq I\leq 19}$ & $%
\oplus $ & $\left\{ \mathbf{\delta }_{a}\right\} _{1\leq a\leq 3}$%
\end{tabular}%
.
\end{equation}%
Then we have:\newline
($\alpha $) \emph{Case of the 2- cycles} $\left\{ C^{I}\right\} _{1\leq
I\leq 19}$ \newline
For the \emph{19} dimensional sub- basis $\left\{ C^{I}\right\} $, the
quaternionic 2- form $\boldsymbol{J}^{ij}$ reads as 
\begin{equation}
\begin{tabular}{llll}
$\boldsymbol{J}^{ij}$ & $=$ & $\mathcal{B}_{NS}\varepsilon ^{ij}+\boldsymbol{%
J}^{\left( ij\right) }$ & ,%
\end{tabular}
\label{qua}
\end{equation}%
and eqs(\ref{jij}) are just the periods, 
\begin{equation}
\begin{tabular}{llllllll}
$\phi ^{ijI}$ & $=$ & $\int_{C^{I}}\boldsymbol{J}^{ij}$ & , & $I$ & $=$ & $%
1,...,19$ & .%
\end{tabular}%
\end{equation}%
In eq(\ref{qua}), $\mathcal{B}_{NS}$ is the NS-NS 2-form B-field and $%
\boldsymbol{J}^{\left( ij\right) }$ the real isotriplet 2-form standing for
the hyperkahler 2-form on K3%
\begin{equation}
\begin{tabular}{llll}
$\boldsymbol{J}^{\left( ij\right) }$ & $=$ & $\left( 
\begin{array}{c}
\Omega ^{\left( 2,0\right) } \\ 
\Omega ^{\left( 1,1\right) } \\ 
\Omega ^{\left( 0,2\right) }%
\end{array}%
\right) $ & ,%
\end{tabular}%
\end{equation}%
where $\Omega ^{\left( 1,1\right) }$ is the usual Kahler 2-form while $%
\Omega ^{\left( 2,0\right) }$ and $\Omega ^{\left( 2,0\right) }$ are the
holomorphic and anti-holomorphic 2-forms on K3.\newline
($\beta $) \emph{Case of the 2- cycles} $\left\{ D^{a}\right\} _{1\leq a\leq
3}$ \newline
For the \emph{3} dimensional sub- basis $\left\{ D^{a}\right\} $, the
quaternionic 2- form $\boldsymbol{J}^{ij}$ reads as%
\begin{equation}
\begin{tabular}{llll}
$\boldsymbol{J}^{ij}$ & $=$ & $\Omega \varepsilon ^{ij}+\mathcal{B}%
_{NS}^{\left( ij\right) }$ & .%
\end{tabular}
\label{qui}
\end{equation}%
where $\Omega =\Omega ^{\left( 1,1\right) }$ is the Kahler 2-form and the
isotriplet $\mathcal{B}_{NS}^{\left( ij\right) }$ as described below.
Indeed, using eqs(\ref{19}-\ref{21}), it follows that the twentieth quartet $%
\phi _{20}^{ij}=\chi ^{ij}=\left( \phi ^{0},\chi ^{\left( ij\right) }\right) 
$ can be written as the periods, 
\begin{equation}
\begin{tabular}{llll}
$\phi ^{0}$ & $=$ & $\int_{B^{20}}\Omega ^{\left( 1,1\right) }$ & , \\ 
$\chi ^{a}$ & $=$ & $\int_{D^{a}}\mathcal{B}_{NS}$ & ,%
\end{tabular}
\label{fx}
\end{equation}%
from which we learn that $\chi ^{a}\sim \chi ^{\left( ij\right) }$ is an
isotriplet as required by the signature of $H_{2}\left( K3,R\right) $. This
property allows to set,%
\begin{equation}
\begin{tabular}{llll}
$\int_{D^{a}}\mathcal{B}_{NS}$ & $=$ & $\int_{\mathrm{C}_{20}}\mathcal{B}%
_{NS}^{a}$ & ,%
\end{tabular}
\label{fy}
\end{equation}%
and think about the \emph{19 + 1 }$=$\emph{20} quartets as $\left( \phi
_{I}^{ij}\right) _{1\leq I\leq 20}$ with the interpretation%
\begin{equation}
\begin{tabular}{llllll}
$\phi ^{ijI}$ & $=$ & $\int_{\mathrm{C}^{I}}\boldsymbol{J}^{ij}$ & , & $%
I=1,...,20$ & ,%
\end{tabular}%
\end{equation}%
where $\left( \mathrm{C}_{I}\right) _{1\leq I\leq 19}$ as in eq(\ref{cd})
and $\int_{\mathrm{C}_{20}}\left( \Omega \varepsilon ^{ij}+\mathcal{B}%
_{NS}^{\left( ij\right) }\right) $ as in eqs(\ref{fx}-\ref{fy}).

\subsection{Deriving the quaternionic potential (\protect\ref{qp})}

\qquad The real \emph{80} fields moduli $\phi _{i}^{jI}=$ $\phi
_{i}^{jI}\left( x\right) $ we have been using are six dimensional real
scalar field variables obeying the reality condition (\ref{re}). These
fields parameterize the coset manifold,%
\begin{equation}
\begin{tabular}{llll}
$\boldsymbol{Q}_{80}$ & $=$ & $\frac{SO\left( 4,20\right) }{SO\left(
4\right) \times SO\left( 20\right) }$ & $.$%
\end{tabular}%
\end{equation}%
The reality condition (\ref{re}) is required by $\mathcal{N}=2$
supersymmetric gauge theory which demands that gauge vector supermultiplets
should be in the (real) adjoint representation of gauge groups (\ref{tra}). 
\newline
As shown by eqs(\ref{g},\ref{gc},\ref{gf}), to deal with the hyper- matter $%
\phi ^{ij}$ it is more convenient to use the complex isodoublet field
variables, 
\begin{equation}
\begin{tabular}{llllllll}
$\mathrm{\phi }^{iA}$ & $=$ & $\left( \mathrm{\phi }^{1A},\mathrm{\phi }%
^{2A}\right) $ & $,$ & $\overline{\mathrm{\phi }}_{iA}$ & $=$ & $\left( 
\overline{\mathrm{\phi }}_{1A},\overline{\mathrm{\phi }}_{2A}\right) $ & .%
\end{tabular}
\label{do}
\end{equation}%
Recall that besides the relation $SO\left( 4\right) \simeq SU\left( 2\right)
\times SU\left( 2\right) $, the switch from the real $\phi _{i}^{k}$ to the
complex coordinates $\mathrm{\phi }^{i}$ has been as well motivated from the
two following: \newline
(\textbf{i}) the wish to exhibit manifestly the $U^{20}\left( 1\right) $
invariance, and \newline
(\textbf{ii}) the objective to use the harmonic superspace method for
building quaternionic metrics.

\subsubsection{Harmonic superspace method}

\qquad Here, we review briefly some useful tools on \emph{HSS} method; in
particular those aspects concerning the superfields $\Phi _{A}^{+}$, that
describe off shell hypermultiplets, the quaternionic potential $\mathcal{L}%
_{int}^{4+}=\mathcal{L}_{int}^{4+}\left( \Phi ^{+},\tilde{\Phi}^{+}\right) $%
, and the way to get the quaternionic metrics $G_{ab}^{IJ}$ in this
superfield theory set up.\newline

\emph{6D} $\mathcal{N}=1$ \emph{harmonic superspace formalism}\newline
Like in the case of \emph{4D} $\mathcal{N}=2$ supersymmetry, the $\mathcal{N}%
=2$ vector representation $\mathcal{V}_{6D}^{\mathcal{N}=2}$ splits into the
sum of a $\mathcal{N}=1$ vector representation $\mathcal{V}_{6D}^{\mathcal{N}%
=1}$ and a hypermultiplet $\mathcal{H}_{6D}^{\mathcal{N}=1}$; eq(\ref{spl}).
This splitting allows here also to use the $\mathcal{N}=1$ superspace
formalism to study hypermultiplet interactions. \newline
In \emph{HSS} with supercoordinates,%
\begin{equation}
\begin{tabular}{llll}
$Z$ & $=$ & $\left\{ \left( x,\theta ^{+},u^{\pm }\right) \text{ };\text{ }%
\theta ^{-}\right\} $ & ,%
\end{tabular}%
\end{equation}%
a generic hypermultiplet $\mathcal{H}_{6D}^{\mathcal{N}=1}$ is described%
\footnote{%
There are two main ways to describe hypermultiplets in terms of HSS\
superfields. One is hermitian and the other is complex as in (\ref{fp}).} by
complex (graded analytic) superfields 
\begin{equation}
\begin{tabular}{llll}
$\Phi _{A}^{+}$ & $=$ & $\Phi _{A}^{+}\left( x,\theta ^{+},u\right) $ & ,%
\end{tabular}
\label{fp}
\end{equation}%
with no dependence in $\theta ^{-}$ and the typical $\theta ^{+}$- expansion 
\begin{equation}
\begin{tabular}{llll}
$\Phi _{A}^{+}$ & $=$ & $\ \ \mathrm{\phi }_{A}^{+}+\theta ^{+\alpha }\theta
^{+\beta }B_{\left[ {\small \alpha \beta }\right] A}^{-}+\theta ^{+4}\Delta
_{A}^{---}$ &  \\ 
&  & $+$ $\ \theta ^{+\alpha }\psi _{\alpha A}+\theta ^{+\alpha }\theta
^{+\beta }\theta ^{+\gamma }\chi _{\left[ {\small \alpha \beta \gamma }%
\right] A}^{--}$ & .%
\end{tabular}
\label{sup}
\end{equation}%
The 6D spinor $\theta =\left( \theta ^{+\alpha }\right) $ is the usual
superspace Grassmann variables. Moreover, each component field 
\begin{equation}
\begin{tabular}{llll}
$F_{A}^{q}$ & $=$ & $F_{A}^{q}\left( x,u\right) \qquad ,\qquad q\in \mathbb{Z%
}$ & ,%
\end{tabular}%
\end{equation}%
of the development (\ref{sup}) is a function of the space time coordinates $%
x=\left( x^{\mu }\right) $ and the harmonic variables $u^{\pm }=\left(
u_{i}^{\pm }\right) $ with the harmonic expansion%
\begin{equation}
\begin{tabular}{llll}
$F_{A}^{q}$ & $=$ & $\sum\limits_{n+q\geq
0}u_{(i_{1}}^{+}...u_{i_{n+q}}^{+}u_{j_{1}}^{-}...u_{j_{n})}^{-}F_{A}^{%
\left( i_{1}...i_{n+q}j_{1}...j_{n}\right) }\left( x\right) $ & .%
\end{tabular}%
\end{equation}%
In particular, we have for the two leading components,%
\begin{equation}
\begin{tabular}{llll}
$\mathrm{\phi }^{+}\left( x,u\right) $ & $=$ & $u_{i}^{+}\mathrm{\phi }%
^{i}\left( x\right) +u_{i}^{+}u_{j}^{+}u_{k}^{-}\mathrm{\phi }^{\left(
ijk\right) }\left( x\right) +...$ & ,%
\end{tabular}
\label{ex}
\end{equation}%
where $\mathrm{\phi }^{i}\left( x\right) $ is precisely the scalar field
given by eq(\ref{do}). \newline
The extra fields $\phi ^{\left( i_{1}....i_{m}\right) }$, $m>2$ of eq(\ref%
{ex}) as well as the following, 
\begin{equation}
\begin{tabular}{llll}
$B_{\left[ {\small \alpha \beta }\right] }^{-}$ & $\sim $ & $B_{\mu
}^{-}\left( x,u\right) $ & , \\ 
$\Delta ^{---}$ & $=$ & $\Delta ^{---}\left( x,u\right) $ & ,%
\end{tabular}
\label{ey}
\end{equation}%
are auxiliary fields required by off shell supersymmetry. They play a
crucial role in the determination of the quaternionic metric we are looking
for. \newline
Notice that the determination of the explicit field expression of these
auxiliary fields in terms of the physical degrees of freedom \textrm{f}$%
^{iA} $ and $\overline{\mathrm{f}}_{iA}$ is one of the main difficult step
in using \emph{HSS} method. \newline
As we will show later on, this difficulty can be overcome in the present
case by help of the $U^{20}\left( 1\right) $ symmetry.\newline

\emph{From the superfield action to the metric}\newline
Generally, the superfield action $\mathcal{S}$, describing hypermultiplet
interactions, reads in rigid harmonic superspace as follows 
\begin{equation}
\begin{tabular}{llll}
$\mathcal{S}$ & $\simeq $ & $\int d^{6}xd^{4}\theta ^{+}\left[ \tilde{\Phi}%
^{+}D^{++}\Phi ^{+}+\mathcal{L}_{int}^{+4}\left( \Phi ^{+},\tilde{\Phi}%
^{+}\right) \right] $ & ,%
\end{tabular}
\label{s}
\end{equation}%
where $D^{++}$ is the harmonic derivative whose basic properties are
collected in the appendix. \newline
The superfield equation of motion of the hypermultiplet $\Phi ^{+}$ reads as%
\begin{equation}
\begin{tabular}{llll}
$\left( D^{++}\Phi ^{+}+\frac{\partial \mathcal{L}_{int}^{+4}}{\partial 
\tilde{\Phi}^{+}}\right) $ & $=$ & $0$ & .%
\end{tabular}
\label{em}
\end{equation}%
Eq(\ref{em}) describes the dynamics of the degrees of freedom \textrm{f}$%
^{iA}$ and $\overline{\mathrm{f}}_{iA}$; but gives also constraint eqs on
the auxiliary fields $\mathrm{\phi }^{\left( i_{1}....i_{m}\right) }$, $B_{%
\left[ {\small \alpha \beta }\right] }^{-}$ and $\Delta ^{---}$. \newline
For example, the equation of motion of the auxiliary field $\Delta ^{---}$
corresponding to the highest $\theta ^{+}$- term in the expansion (\ref{ex}%
), reads as follows. 
\begin{equation}
\begin{tabular}{llll}
$\left( u^{+i}\frac{\partial \mathrm{\phi }^{+}}{\partial u^{-i}}+\frac{%
\partial \mathcal{L}_{int}^{+4}}{\partial \mathrm{\phi }^{+}}\right) $ & $=$
& $0$ & ,%
\end{tabular}
\label{uu}
\end{equation}%
This relation is a constraint equation that fix the auxiliary fields of eq(%
\ref{ex}) as follows,%
\begin{equation}
\begin{tabular}{llll}
${\normalsize \phi }^{+}$ & $=$ & $\mathrm{\phi }^{+}\left( \mathrm{f}^{+A},%
\mathrm{f}^{-A}\right) $ & ,%
\end{tabular}
\label{fia}
\end{equation}%
where 
\begin{equation}
\begin{tabular}{llllll}
$\mathrm{f}^{\pm A}\left( x,u\right) $ & $=$ & $\sum\limits_{i=1}^{2}u_{i}^{%
\pm }\mathrm{f}^{iA}\left( x\right) $ & , & $A=1,...,n$ & .%
\end{tabular}%
\end{equation}%
A quite similar relation to eq(\ref{uu}), can be also written down for $%
B_{\mu }^{-}$; and its solution gives the expression of $B_{\mu }^{-}$ in
terms of the scalar moduli $\mathrm{f}^{\pm A}$; i.e%
\begin{equation}
\begin{tabular}{llll}
$B_{\mu }^{-}$ & $=$ & $B_{\mu }^{-}\left( \mathrm{f}^{+A},\mathrm{f}%
^{-A}\right) $ & .%
\end{tabular}
\label{bia}
\end{equation}%
With eqs(\ref{fia}-\ref{bia}) in mind; and integrating eq(\ref{s}) with
respect to the Grassmann variables $\theta ^{+\alpha }$, we can bring the
superfield action to the following remarkable form,%
\begin{equation}
\begin{tabular}{llll}
$\mathcal{S}$ & $=$ & $\int d^{6}xdu\left( \overline{B}_{\mu }^{-}\partial
^{\mu }\mathrm{\phi }^{+}-B_{\mu }^{-}\partial ^{\mu }\overline{\mathrm{\phi 
}}^{+}\right) $ & .%
\end{tabular}
\label{bf}
\end{equation}%
Substituting $\mathrm{\phi }^{+}$ and $B_{\mu }^{-}$ by their expression (%
\ref{fia}-\ref{bia}) and integrating with respect to the harmonic variables $%
u^{\pm i}$, we can further bring (\ref{bf}) to the form%
\begin{equation}
\mathcal{S}=\int d^{6}x\left( 2h_{iA}^{jB}\partial _{\mu }\mathrm{f}%
^{iA}\partial ^{\mu }\overline{\mathrm{f}}_{jB}+\overline{g}_{iA,jB}\partial
_{\mu }\mathrm{f}^{iA}\partial ^{\mu }\mathrm{f}^{jB}+g^{iA,jB}\partial
_{\mu }\overline{\mathrm{f}}_{iA}\partial _{\mu }\overline{\mathrm{f}}%
_{jB}\right) ,  \label{ss}
\end{equation}%
from which we read the expression of $\overline{g}_{iA,jB}$, $g^{iA,jB}$ and 
$h_{iA}^{jB}$ associated with eq(\ref{qp}).\newline
Notice that bringing the action $\mathcal{S}$ from its expression (\ref{s})
into the form (\ref{ss}) is in fact a very complicated task; except for some
special situations where there are symmetries. As we will see for the case
at hand, it is possible to put (\ref{s}) with eq(\ref{qp}) into the form (%
\ref{ss}); thanks to the $U^{20}\left( 1\right) $ invariance that we want to
study below.

\subsubsection{Complex fields and $U^{20}\left( 1\right) $ symmetry}

\qquad First we study the field duality mapping the real field coordinates $%
\phi _{j}^{i}$ into the complex ones $\mathrm{\phi }^{i}$ and $\overline{%
\mathrm{\phi }}_{i}$. Then, we describe the $U^{20}\left( 1\right) $
invariance of eq(\ref{gi}).\newline

\emph{From real }$\phi _{j}^{i}$\emph{\ to complex }$\mathrm{\phi }^{i}$%
\emph{\ fields }\newline
A typical field change\textrm{\ }that relates the four real variables,
described by the SU$\left( 2\right) $ rank 2- tensor $\phi _{j}^{i},$ to the
complex isodoublets $\left( \mathrm{f}^{i}\right) $ and $\left( \overline{%
\mathrm{f}}_{i}\right) $ is given by,%
\begin{equation}
\begin{tabular}{llllllll}
$\phi _{j}^{i}$ & $=$ & $\overline{\mathrm{f}}_{j}\mathrm{f}^{i}$ & , & $%
\overline{\left( \phi _{j}^{i}\right) }$ & $=$ & $\phi _{i}^{j}$ & .%
\end{tabular}
\label{ch}
\end{equation}%
The complex scalar fields $\mathrm{f}^{i}=\left( \mathrm{f}^{1},\mathrm{f}%
^{2}\right) $ capture two complex degrees of freedom and can be expressed in
terms of $\phi _{j}^{i}$ as follows%
\begin{equation}
\begin{tabular}{llll}
$\mathrm{f}^{i}$ & $=$ & $\sum\limits_{k=1}^{2}\mathrm{\upsilon }^{k}\phi
_{k}^{i}$ & $,$ \\ 
$\overline{\mathrm{f}}_{i}$ & $=$ & $\sum\limits_{k=1}^{2}\overline{\mathrm{%
\upsilon }}_{k}\phi _{i}^{k}$ & $.$%
\end{tabular}
\label{fu}
\end{equation}%
Multiplying both sides of eq(\ref{ch}) by $\mathrm{f}^{j}$ and by using the
following relations,%
\begin{equation}
\begin{tabular}{llll}
$\phi _{j}^{i}$ & $=$ & $\frac{1}{2}\phi ^{0}\delta _{j}^{i}+\frac{1}{2}%
\sum\limits_{k=1}^{2}\varepsilon _{jk}\phi ^{\left( ik\right) }$ & ,%
\end{tabular}%
\end{equation}%
and%
\begin{equation}
\begin{tabular}{llll}
$\phi ^{0}$ & $=$ & $\mathrm{f}^{i}\overline{\mathrm{f}}_{i}\equiv \mathrm{f}%
\overline{\mathrm{f}}$ & , \\ 
$\phi ^{\left( ik\right) }$ & $=$ & $\mathrm{f}^{i}\overline{\mathrm{f}}^{k}+%
\mathrm{f}^{k}\overline{\mathrm{f}}^{i}$ & ,%
\end{tabular}%
\end{equation}%
we obtain eq(\ref{fu}) with%
\begin{equation}
\begin{tabular}{llll}
$\mathrm{\upsilon }^{k}$ & $=$ & $\frac{\mathrm{f}^{k}}{\mathrm{f}\overline{%
\mathrm{f}}}=\frac{\mathrm{f}^{k}}{\varphi ^{0}}$ & , \\ 
$\overline{\mathrm{\upsilon }}_{k}$ & $=$ & $\frac{\overline{\mathrm{f}}_{k}%
}{\mathrm{f}\overline{\mathrm{f}}}=\frac{\overline{\mathrm{f}}_{k}}{\varphi
^{0}}$ & ,%
\end{tabular}%
\end{equation}%
satisfying the identities%
\begin{equation}
\begin{tabular}{llllllll}
$\sum\limits_{k=1}^{2}\overline{\mathrm{\upsilon }}_{k}\mathrm{\upsilon }%
^{k} $ & $=$ & $1$ & , &  &  &  &  \\ 
$\sum\limits_{k,l=1}^{2}\varepsilon _{kl}\mathrm{\upsilon }^{l}\mathrm{%
\upsilon }^{k}$ & $=$ & $0$ & , & $\sum\limits_{k,l=1}^{2}\varepsilon ^{kl}%
\overline{\mathrm{\upsilon }}_{k}\overline{\mathrm{\upsilon }}_{l}$ & $=$ & $%
0$ & .%
\end{tabular}%
\end{equation}

\emph{Implementing the} \emph{U}$^{20}\left( 1\right) $ \emph{gauge} \emph{%
symmetry}\newline
The above field variable change is remarkable and exhibits the following
properties:\newline
(\textbf{i}) Eq(\ref{ch}) has a manifest $U\left( 1\right) $ abelian
symmetry,%
\begin{equation}
\begin{tabular}{llllll}
$\mathrm{f}^{k}$ & $\rightarrow $ & $\phi ^{k}$ & $=$ & e$^{i\vartheta }%
\mathrm{f}^{k}$ & , \\ 
$\overline{\mathrm{f}}_{k}$ & $\rightarrow $ & $\phi _{k}$ & $=$ & e$%
^{-i\vartheta }\overline{\mathrm{f}}_{k}$ & ,%
\end{tabular}
\label{gs}
\end{equation}%
where the local real function $\vartheta $ is the gauge parameter of the $%
U\left( 1\right) $ invariance.\newline
This $U\left( 1\right) $ symmetry tells us that the fields $\mathrm{f}^{k}$
involved in the duality relation (\ref{ch}) are not uniquely defined. Under
the abelian gauge transformation (\ref{gs}), we have 
\begin{equation}
\begin{tabular}{llllll}
$\phi _{j}^{i}$ & $=$ & $\overline{\mathrm{f}}_{j}\mathrm{f}^{i}$ & $=$ & $%
\overline{\mathrm{\phi }}_{j}\mathrm{\phi }^{i}$ & ,%
\end{tabular}%
\end{equation}%
showing that $\phi _{j}^{i}$ can be interpreted as conserved quantity; that
is a conserved Noether current in the field theory set up. This conserved
quantity is precisely the one given by eq(\ref{313}).\newline
(\textbf{ii}) The duality relation (\ref{ch}) can be generalized as follows,%
\begin{equation}
\begin{tabular}{llllll}
$\phi _{k}^{iI}$ & $=$ & $\overline{\mathrm{f}}_{k}T^{I}\mathrm{f}^{i}$ & $=$
& $\tsum\limits_{A,B=1}^{n}\overline{\mathrm{f}}_{kB}\left( T^{I}\right)
_{A}^{B}\mathrm{f}^{iA}$ & ,%
\end{tabular}
\label{chh}
\end{equation}%
where now the matter fields $\mathrm{f}^{iA}$ are in the $\left( \underline{2%
},\underline{n}\right) $ bi-fundamental of $SU\left( 2\right) \times U\left(
n\right) $. For the case $n=20$, we have, 
\begin{equation}
\begin{tabular}{llllll}
$\mathrm{f}^{iA}$ & $\simeq $ & $\left( \underline{2},\underline{20}\right) $
& $\in $ & $SU\left( 2\right) \times U\left( 20\right) $ & ,%
\end{tabular}%
\end{equation}%
where the $20\times 20$ commuting matrices $T^{I}$ are as in eqs(\ref{t}).
They are the commuting Cartan generators of the $U\left( 20\right) $ unitary
group.\newline
Notice that the duality relation (\ref{chh}) has the following manifest
abelian $U^{20}\left( 1\right) $ gauge symmetry%
\begin{equation}
\begin{tabular}{llllll}
$\mathrm{f}^{kA}$ & $\rightarrow $ & $\phi ^{kA}$ & $=$ & $\left(
e^{i\vartheta }\mathrm{f}^{k}\right) ^{A}$ & , \\ 
$\overline{\mathrm{f}}_{kA}$ & $\rightarrow $ & $\overline{\phi }_{kA}$ & $=$
& $\left( e^{-i\vartheta }\overline{\mathrm{f}}_{k}\right) _{A}$ & .%
\end{tabular}%
\end{equation}%
These transformations read more explicitly like,%
\begin{equation}
\begin{tabular}{llllll}
$\mathrm{f}^{kA}$ & $\rightarrow $ & $\phi ^{kA}$ & $=$ & $%
\tsum\limits_{C=1}^{20}\left( e^{i\vartheta }\right) _{C}^{A}\mathrm{f}^{kC}$
& , \\ 
$\overline{\mathrm{f}}_{kB}$ & $\rightarrow $ & $\overline{\phi }_{kB}$ & $=$
& $\tsum\limits_{D=1}^{20}\left( e^{-i\vartheta }\right) _{B}^{D}\overline{%
\mathrm{f}}_{kD}$ & ,%
\end{tabular}%
\end{equation}%
with $\vartheta $ given by the expansion 
\begin{equation}
\begin{tabular}{llll}
$\vartheta $ & $=$ & $\tsum\limits_{I=1}^{20}\vartheta _{I}T^{I}$ & ,%
\end{tabular}%
\end{equation}%
where the $\vartheta _{I}^{\prime }$s are the \emph{20} gauge parameters of
the $U^{20}\left( 1\right) $ abelian invariance. It is not difficult to
check that we have the identity, 
\begin{equation}
\begin{tabular}{llllll}
$\varphi _{k}^{iI}$ & $=$ & $\overline{\mathrm{f}}_{k}T^{I}\mathrm{f}^{i}$ & 
$=$ & $\overline{\mathrm{\phi }}_{k}T^{I}\mathrm{\phi }^{i}$ & ,%
\end{tabular}
\label{fik}
\end{equation}%
showing that $\phi _{k}^{iI}$ is a conserved quantity; thanks to eqs(\ref{t}%
). \newline
In \emph{HSS}\ formalism, the relation (\ref{fik}) corresponds to the
leading term of the $\theta ^{+}$- expansion of the \emph{HSS} relation $%
H^{++I}=\mathrm{\tilde{\Phi}}^{+}T^{I}\mathrm{\Phi }^{+}$ as given by eq(\ref%
{312}). These conserved quantities ($D^{++}H^{++I}=0$) play a determinant
role in the solving the underlying constraint eqs that lead to the
computation of the metric of the moduli space $Q_{80}$.

(\textbf{c}) \emph{Comments}\newline
Below, we make three comments regarding the use of the complex field
variables $\left( \mathrm{f}^{kA},\overline{\mathrm{f}}_{kA}\right) $ rather
than the $\phi _{k}^{iI}$ real ones.\newline
(\textbf{i}) the complex fields $\mathrm{f}^{kA}$ are in the bi-fundamental
of $SU\left( 2\right) \times U\left( 20\right) $. They parameterize the
complex \emph{40} dimensional coset manifold, 
\begin{equation}
\begin{tabular}{llll}
$\boldsymbol{H}_{20}$ & $=$ & $\frac{SU\left( 2,20\right) }{SU\left(
2\right) \times U\left( 20\right) }$ & $,$%
\end{tabular}%
\end{equation}%
which is contained in $\frac{SO\left( 4,20\right) }{SO\left( 4\right) \times
SO\left( 20\right) }$. The manifold $\boldsymbol{H}_{20}$ has a richer
isotropy symmetry.\newline
The isosinglets $\phi _{I}^{0}$ and isotriplets $\phi _{I}^{\left( kl\right)
}$ are now given by%
\begin{equation}
\begin{tabular}{llll}
$\phi _{I}^{0}$ & $=$ & $\overline{\mathrm{\phi }}_{k}T_{I}\mathrm{\phi }%
^{k} $ & , \\ 
$\phi _{I}^{\left( kl\right) }$ & $=$ & $\overline{\mathrm{\phi }}^{k}T_{I}%
\mathrm{\phi }^{l}+\overline{\mathrm{\phi }}^{l}T_{I}\mathrm{\phi }^{k}$ & .%
\end{tabular}
\label{ff}
\end{equation}%
By using the quaternionic form $\boldsymbol{J}^{kl}$ introduced previously,
we can also rewrite the above relations collectively like,%
\begin{equation}
\begin{tabular}{llllll}
$\overline{\mathrm{\phi }}^{k}T^{I}\mathrm{\phi }^{l}$ & $=$ & $\int_{C^{I}}%
\boldsymbol{J}^{kl}$ & $,$ & $I=1,...,20$ & .%
\end{tabular}
\label{lo}
\end{equation}%
(\textbf{ii}) Eqs(\ref{ff}) appear in the \emph{HSS} method as the lowest
component of the $\theta ^{+}$- expansion of the analytic superfield 
\begin{equation}
\begin{tabular}{llll}
$H_{I}^{++}$ & $=$ & $\tilde{\Phi}^{+}T_{I}\Phi ^{+}$ & ,%
\end{tabular}
\label{vf}
\end{equation}%
that is 
\begin{equation}
\begin{tabular}{llll}
$\sum\limits_{i,j=1}^{2}u_{k}^{+}u_{l}^{-}\left( \overline{\mathrm{\phi }}%
^{k}T^{I}\mathrm{\phi }^{l}\right) $ & $=$ & $\sum%
\limits_{i,j=1}^{2}u_{k}^{+}u_{l}^{-}\left[ \tilde{\Phi}^{k}T_{I}\Phi ^{l}%
\right] _{\theta =0}$ & .%
\end{tabular}%
\end{equation}%
Eq(\ref{vf}) obeys $D^{++}H_{I}^{++}=0$ and solved like 
\begin{equation}
\begin{tabular}{llll}
$H^{++I}$ & $=$ & $\sum\limits_{i,j=1}^{2}u_{i}^{+}u_{j}^{+}H^{ijI}$ & .%
\end{tabular}%
\end{equation}%
They may be interpreted "periods" of some \emph{10D} superspace 2-form $%
\mathcal{J}^{++}=u_{i}^{+}u_{j}^{+}\mathcal{J}^{ij}$ as%
\begin{equation}
\begin{tabular}{llllll}
$H^{ijI}$ & $=$ & $\int_{C^{I}}\mathcal{J}^{ij}$ & , & $I=1,...,20$ & ,%
\end{tabular}
\label{vq}
\end{equation}%
or equivalently%
\begin{equation}
\begin{tabular}{llll}
$H^{++I}$ & $=$ & $\int_{C^{I}}\mathcal{J}^{++}$ & ,%
\end{tabular}
\label{vp}
\end{equation}%
with lowest $\theta ^{+}$- component as in eq(\ref{lo}).\newline
(\textbf{iii}) Using eq(\ref{vp}), we can compute the \emph{HSS} potential
that describe the moduli space of 10D type IIA superstring on K3%
\begin{equation}
\begin{tabular}{llll}
$\mathcal{L}_{int}^{+4}$ & $\simeq $ & $\int_{K3}\mathcal{J}^{++}\wedge 
\mathcal{J}^{++}$ & .%
\end{tabular}%
\end{equation}%
Upon integration, we get precisely the relation (\ref{qp}). \newline
With these tools at hand, we are in position to compute the explicit field
expression of the quaternionic metric (\ref{ss}).

\section{Quaternionic metric}

\qquad In this section, we use the complex coordinates $\left( \mathrm{f}%
^{iA},\overline{\mathrm{f}}_{iA}\right) $ and derive the explicit expression
of the quaternionic metric $G_{ab}^{IJ}=G_{ab}^{IJ}\left( \mathrm{f,}%
\overline{\mathrm{f}}\right) $ of the scalar manifold of the \emph{6D} $%
\mathcal{N}=2$ supergravity.\newline
Since the result we give here is valid for the real \emph{4n} dimensional
manifolds%
\begin{equation}
\begin{tabular}{llllllll}
$\boldsymbol{H}_{2n}$ & $=$ & $\frac{SU\left( 2,n\right) }{SU\left( 2\right)
\times U\left( n\right) }$ & , & $\dim \boldsymbol{H}_{2n}$ & $=$ & $4n$ & .%
\end{tabular}
\label{hn}
\end{equation}%
and in order to be as much as general, we will proceed as follow:\newline
After, showing how the dilaton factorizes, we focus on the derivation of the
quaternionic metric of the scalar manifold $\boldsymbol{H}_{2}$ for the case
of \emph{6D} $\mathcal{N}=2$ supergravity multiplet coupled to a $n=1$
Maxwell supermultiplet. There, the scalar manifold $\boldsymbol{H}_{2}$ is
given by,%
\begin{equation}
\begin{tabular}{llll}
$\boldsymbol{H}_{2}$ & $=$ & $\frac{SU\left( 2,1\right) }{SU\left( 2\right)
\times U\left( 1\right) }$ & ,%
\end{tabular}
\label{2}
\end{equation}%
and corresponds to the real 4- dimensional Taub-NUT model.\newline
Then, we consider the computation of the metric for the generic real $4n$
dimensional scalar manifold (\ref{hn}). The \emph{6D} $\mathcal{N}=2$
supergravity embedded in \emph{10D} type IIA superstring on K3 is obtained
by setting $n=20$.

\subsection{Taub-NUT geometry}

\qquad This geometry concerns the scalar manifold $\boldsymbol{H}_{2}$
involved in the the moduli space 
\begin{equation}
\begin{tabular}{llll}
$SO\left( 1,1\right) \times \boldsymbol{H}_{2}$ & $\subset $ & $SO\left(
1,1\right) \times \frac{SO\left( 4,1\right) }{SO\left( 4\right) }$ & ,%
\end{tabular}%
\end{equation}%
of the $\mathcal{N}=2$ gravity supermultiplet coupled to one Maxwell
supermultiplet; i.e: $n=1$.\newline

\emph{Field theory set up}\newline
In the real field coordinate frame $\left\{ \sigma ,\phi ^{a}\right\} $ of
the moduli space $SO\left( 1,1\right) \times \frac{SO\left( 4,1\right) }{%
SO\left( 4\right) }$, the component field action $\mathcal{S}$ describing
the underlying non linear sigma model reads as follows 
\begin{equation}
\begin{tabular}{lllllll}
$\mathcal{S}$ & $=$ & $\int d^{6}x\sqrt{-\det \text{g}}\left( \mathcal{R}%
-g^{\mu \nu }\partial _{\mu }\sigma \partial _{\nu }\sigma \right) $ & $+$ & 
$\mathcal{S}_{1}$ & $+$ \ ... & .%
\end{tabular}%
\end{equation}%
Here $\mathcal{R}$ is the 6D space time scalar curvature, $\sigma $ the
dilaton associated with the factor $SO\left( 1,1\right) $ and%
\begin{equation}
\begin{tabular}{lllll}
$\mathcal{S}_{1}$ & $=$ & $\int d^{6}x\sqrt{-\det \text{g}}\left[ e^{-\sigma
}g^{\mu \nu }\left( \partial _{\mu }\phi ^{a}\partial _{\nu }\phi
^{b}G_{ab}\right) \right] $ & $+$ $...$ & .%
\end{tabular}
\label{s1}
\end{equation}%
The dots in the above relations stand for the extra terms required by
supersymmetry and $G_{ab}=G_{ab}\left( \phi \right) $ is the metric of the
factor $\frac{SO\left( 4,1\right) }{SO\left( 4\right) }$.\newline
Below, we focus on the study of the scalar field contribution in $\mathcal{S}%
_{1}$ by using the complex fields $\mathrm{f}^{i}$ and $\overline{\mathrm{f}}%
_{k}$ variables given by the field coordinates change 
\begin{equation}
\begin{tabular}{llllllll}
$\phi _{k}^{i}$ & $=$ & $\overline{\mathrm{f}}_{k}\mathrm{f}^{i}$ & , & $%
\overline{\left( \phi _{k}^{i}\right) }$ & $=$ & $\phi _{i}^{k}$ & ,%
\end{tabular}%
\end{equation}%
together with the $U\left( 1\right) $ gauge symmetry (\ref{gs}). The complex
isodoublet $\mathrm{f}^{i}$ parameterize the manifold $\boldsymbol{H}_{2}$
given by eq(\ref{1}). \newline
To get the expression of the scalar field part $\mathcal{S}_{1}^{scalar}$ of
the field action (\ref{s1}), we first substitute,%
\begin{equation}
\begin{tabular}{llll}
$\partial _{\nu }\phi _{k}^{i}$ & $=$ & $\overline{\mathrm{f}}_{k}\partial
_{\nu }\mathrm{f}^{i}+\mathrm{f}^{i}\partial _{\nu }\overline{\mathrm{f}}%
_{k} $ & ,%
\end{tabular}%
\end{equation}%
which allows to bring $\mathcal{S}_{1}^{scalar}$ to,%
\begin{equation}
\begin{tabular}{llll}
$\mathcal{S}_{1}^{scalar}$ & $\simeq $ & $\int d^{6}x\sqrt{-\det \text{g}}%
L_{1}$ & ,%
\end{tabular}
\label{ls}
\end{equation}%
with%
\begin{equation}
\begin{tabular}{llll}
$L_{1}$ & $=$ & $\left( B_{\mu }^{i}\partial ^{\mu }\overline{\mathrm{f}}%
_{i}+\overline{B}_{\mu i}\partial ^{\mu }\mathrm{f}^{i}\right) $ & ,%
\end{tabular}
\label{sl}
\end{equation}%
where we have set $d\sigma =0$. In this relation, the factor $B_{\mu }^{i}$
is a function of the physical degrees of freedom $\mathrm{f}^{i},$ 
\begin{equation}
\begin{tabular}{llllllll}
$B_{\mu }^{i}$ & $=$ & $B_{\mu }^{i}\left( \mathrm{f,}\overline{\mathrm{f}}%
\right) $ & , & $\overline{B}_{\mu i}$ & $=$ & $\overline{B}_{\mu i}\left( 
\mathrm{f,}\overline{\mathrm{f}}\right) $ & .%
\end{tabular}%
\end{equation}%
The $B_{\mu }^{i}$ captures the scalar fields coupling and should be
compared with eq(\ref{bia}). Notice that the Lagrangian (\ref{ls}-\ref{sl})
has the same structure as (\ref{bf}). This property is just the
manifestation of the fact that $L_{1}$ is nothing but the bosonic part of
the \emph{HSS}\ Lagrangian, 
\begin{equation}
\begin{tabular}{llll}
$\mathcal{L}_{1}^{4+}$ & $=$ & $\mathrm{\tilde{\Phi}}^{+}D^{++}\mathrm{\Phi }%
^{+}-\frac{\lambda }{2}\left( \mathrm{\tilde{\Phi}}^{+}\mathrm{\Phi }%
^{+}\right) ^{2}$ & ,%
\end{tabular}
\label{ltn}
\end{equation}%
where $\mathrm{\Phi }^{+}$ as in eq(\ref{sup}), $\lambda $ is a real
coupling constant and $\left( \mathrm{\tilde{\Phi}}^{+}\mathrm{\Phi }%
^{+}\right) ^{2}$ is the Tub-NUT potential (\ref{tn}).\newline
The \emph{HSS} Lagrangian density (\ref{ltn}) has the $U\left( 1\right) $
symmetry%
\begin{equation}
\begin{tabular}{llllll}
$\mathrm{\Phi }^{+}$ & $\qquad \rightarrow \qquad $ & $\mathrm{\Phi }%
^{+^{\prime }}$ & $=$ & $e^{i\vartheta }\mathrm{\Phi }^{+}$ & ,%
\end{tabular}%
\end{equation}%
with group parameter $\vartheta $. This symmetry should be associated with
eq(\ref{gs}). \newline
Moreover, using covariance under the $SU\left( 2\right) \times U\left(
1\right) $ isotropy symmetry of the scalar manifold, we can also put $B_{\mu
}^{i}$ and $\overline{B}_{\mu i}$ in the form,%
\begin{equation}
\begin{tabular}{llll}
$B_{\mu }^{i}$ & $=$ & $h_{k}^{i}\partial _{\mu }\mathrm{f}%
^{k}+g^{ik}\partial _{\mu }\overline{\mathrm{f}}_{k}$ & , \\ 
$\overline{B}_{\mu i}$ & $=$ & $h_{i}^{k}\partial _{\mu }\overline{\mathrm{f}%
}_{k}+\overline{g}_{ik}\partial _{\mu }\overline{\mathrm{f}}^{k}$ & ,%
\end{tabular}
\label{bb}
\end{equation}%
with 
\begin{equation}
\begin{tabular}{llllllll}
$h_{k}^{i}$ & $=$ & $h_{k}^{i}\left( \mathrm{f,}\overline{\mathrm{f}}\right) 
$ & \qquad ,\qquad & $h_{k}^{i}$ & $=$ & $\overline{\left( h_{i}^{k}\right) }
$ & , \\ 
$g^{ik}$ & $=$ & $g^{ik}\left( \mathrm{f,}\overline{\mathrm{f}}\right) $ & 
\qquad ,\qquad & $\overline{g}_{ik}$ & $=$ & $\overline{\left( g^{ik}\right) 
}$ & .%
\end{tabular}%
\end{equation}%
The same covariance argument allows as well to factorize the metric
components $h_{k}^{i}$, $g^{ik}$ and $\overline{g}_{ik}$ like,%
\begin{equation}
\begin{tabular}{llll}
$h_{k}^{i}$ & $=$ & $\delta _{k}^{i}\left( 1+\xi \right) +\alpha \overline{%
\mathrm{f}}_{k}\mathrm{f}^{i}$ & , \\ 
$g^{ik}$ & $=$ & $\beta \mathrm{f}^{i}\mathrm{f}^{j}$ & , \\ 
$\overline{g}_{ik}$ & $=$ & $\overline{\beta }\overline{\mathrm{f}}_{k}%
\overline{\mathrm{f}}_{i}$ & ,%
\end{tabular}%
\end{equation}%
with,%
\begin{equation}
\begin{tabular}{llllllll}
$\alpha $ & $=$ & $\alpha \left( \rho \right) $ & \qquad ,\qquad & $%
\overline{\left( \alpha \right) }$ & $=$ & $\alpha $ & , \\ 
$\xi $ & $=$ & $\xi \left( \rho \right) $ & \qquad ,\qquad & $\overline{%
\left( \xi \right) }$ & $=$ & $\xi $ & , \\ 
$\beta $ & $=$ & $\beta \left( \rho \right) $ & \qquad ,\qquad & $\overline{%
\left( \beta \right) }$ & $=$ & $\overline{\beta }$ & , \\ 
$\rho $ & $=$ & $\lambda \mathrm{f}\overline{\mathrm{f}}$ & \qquad ,\qquad & 
$\rho $ & $=$ & $\mathrm{f}^{i}\overline{\mathrm{f}}_{i}$ & .%
\end{tabular}
\label{al}
\end{equation}%
The coupling constant $\lambda $ is same as above and may be also
interpreted, in the \emph{6D} black hole physics, as the area of the $%
AdS_{2}\times S^{4}$ near horizon geometry.\newline
Substituting (\ref{bb}) back into (\ref{sl}), we obtain%
\begin{equation}
\begin{tabular}{llll}
$L_{1}$ & $=$ & $\left( 2h_{i}^{j}\partial _{\mu }\mathrm{f}^{i}\partial
^{\mu }\overline{\mathrm{f}}_{j}+\overline{g}_{ij}\partial _{\mu }\mathrm{f}%
^{i}\partial ^{\mu }\mathrm{f}^{j}+g^{ij}\partial _{\mu }\overline{\mathrm{f}%
}_{i}\partial _{\mu }\overline{\mathrm{f}}_{j}\right) $ & ,%
\end{tabular}%
\end{equation}%
which should be compared with eq(\ref{ss}). \newline
On the other hand, following \textrm{\cite{IS,ShS}}, the integration of (\ref%
{ltn}) with respect to the Grassmann variables $\theta ^{+\alpha }$ and the
harmonic variables $u_{i}^{\pm }$ leads to the following,%
\begin{equation}
\begin{tabular}{llll}
$h_{i}^{j}$ & $=$ & $\delta _{i}^{j}\left( 1+\lambda \mathrm{f}\overline{%
\mathrm{f}}\right) -\frac{\lambda }{2}\frac{\left( 2+\lambda \mathrm{f}%
\overline{\mathrm{f}}\right) }{\left( 1+\lambda \mathrm{f}\overline{\mathrm{f%
}}\right) }\mathrm{f}^{j}\overline{\mathrm{f}}_{i}$ & , \\ 
$g_{ij}$ & $=$ & $\frac{\lambda }{2}\frac{\left( 2+\lambda \mathrm{f}%
\overline{\mathrm{f}}\right) }{\left( 1+\lambda \mathrm{f}\overline{\mathrm{f%
}}\right) }\overline{\mathrm{f}}_{i}\overline{\mathrm{f}}_{j}$ & , \\ 
$\overline{g}^{_{ij}}$ & $=$ & $\frac{\lambda }{2}\frac{\left( 2+\lambda 
\mathrm{f}\overline{\mathrm{f}}\right) }{\left( 1+\lambda \mathrm{f}%
\overline{\mathrm{f}}\right) }\mathrm{f}^{i}\mathrm{f}^{j}$ & .%
\end{tabular}%
\end{equation}%
From these relations, we can easily read the explicit field expressions of
the functions $\alpha ,$ $\xi $ and $\beta $ given by eqs(\ref{al}). \newline
Now we turn to derive the quaternionic metric for the generic models (\ref%
{hn}).

\subsection{Generic quaternionic metric}

\qquad In the generic case where the \emph{6D} $N=2$ gravity supermultiplet
is coupled to $n$ Maxwell multiplets, the \emph{HSS} potential $\mathcal{L}%
_{n}^{4+}$ has the structure (\ref{qp}). The superspace Lagrangian density
reads then as follows:%
\begin{equation}
\begin{tabular}{llll}
$\mathcal{L}_{n}^{4+}$ & $=$ & $\int d^{4}\theta ^{+}du\left[ \mathrm{\tilde{%
\Phi}}^{+}D^{++}\mathrm{\Phi }^{+}-\frac{\lambda }{2}\left( \mathrm{\tilde{%
\Phi}}^{+}T^{I}\mathrm{\Phi }^{+}\right) d_{IJ}\left( \mathrm{\tilde{\Phi}}%
^{+}T^{J}\mathrm{\Phi }^{+}\right) \right] $ & .%
\end{tabular}
\label{ln}
\end{equation}%
In this relation the $n\times n$ hermitian matrices $T^{I}$ are the $%
U^{n}\left( 1\right) $ Cartan generators of the group $U\left( n\right) $;
and the symmetric matrix $d_{IJ}$ is a coupling constant matrix which, for
the case $n=20$, has an interpretation in terms of intersections of 2-
cycles of K3.\newline
The equations of motion following from $\mathcal{L}_{n}^{4+}$ read as%
\begin{equation}
\begin{tabular}{llll}
$\left[ D^{++}-\lambda \left( \mathrm{\tilde{\Phi}}^{+}T^{J}\mathrm{\Phi }%
^{+}\right) d_{IJ}T^{I}\right] \mathrm{\Phi }^{+}$ & $=$ & $0$ & , \\ 
$\left[ D^{++}+\lambda \left( \mathrm{\tilde{\Phi}}^{+}T^{J}\mathrm{\Phi }%
^{+}\right) d_{IJ}T^{I}\right] \mathrm{\tilde{\Phi}}^{+}$ & $=$ & $0$ & .%
\end{tabular}%
\end{equation}%
The \emph{HSS} Lagrangian density (\ref{ln}) and the equations of motion are
invariant under the $U^{n}\left( 1\right) $ gauge symmetry,%
\begin{equation}
\begin{tabular}{llllllll}
$\mathrm{\Phi }_{A}^{+\prime }$ & $=$ & $\left( e^{i\Lambda }\right) _{A}^{B}%
\mathrm{\Phi }_{B}^{+}$ & , & $\Lambda $ & $=$ & $\tsum\limits_{I=1}^{n}%
\Lambda _{I}T^{I}$ & .%
\end{tabular}%
\end{equation}%
The conserved \emph{HSS}\ Noether currents corresponding to the gauge
parameters $\Lambda _{I}$ are precisely 
\begin{equation}
\begin{tabular}{llll}
$H_{I}^{++}$ & $=$ & $\mathrm{\tilde{\Phi}}^{+}T_{I}\mathrm{\Phi }^{+}$ & ,%
\end{tabular}%
\end{equation}%
and obey the \emph{HSS} conservation laws $D^{++}H_{I}^{++}=0$. \newline
Moreover, performing the integration of eq(\ref{ln}) with respect to the
Grassmann variables $\theta ^{+}$ and $\overline{\theta }^{+}$, we obtain
the following,%
\begin{equation}
\begin{tabular}{llll}
$\mathcal{L}_{n}$ & $=$ & $\frac{1}{2}\int du\left( B_{\mu }^{-A}\partial
^{\mu }\tilde{\phi}_{A}^{+}-\tilde{B}_{\mu A}^{-}\partial ^{\mu }\phi
^{+A}\right) $ & .%
\end{tabular}
\label{bfa}
\end{equation}%
To put $\mathcal{L}_{n}$ in the form,%
\begin{equation*}
\begin{tabular}{llll}
$\mathcal{L}_{n}$ & $=$ & $\frac{1}{2}\left( 2h_{iA}^{jB}\partial _{\mu }%
\mathrm{f}^{iA}\partial ^{\mu }\overline{\mathrm{f}}_{jB}+\overline{g}%
_{iAjB}\partial _{\mu }\mathrm{f}^{iA}\partial ^{\mu }\mathrm{f}%
^{jB}+g^{iAjB}\partial _{\mu }\overline{\mathrm{f}}_{iA}\partial ^{\mu }%
\overline{\mathrm{f}}_{jB}\right) $ & ,%
\end{tabular}%
\end{equation*}%
we have to perform the two following steps:\newline
(\textbf{1}) determine the explicit field dependence of $\mathrm{\phi }%
_{A}^{+}$ and $B_{\mu A}^{-}$ in terms of the physical $\mathrm{f}_{A}^{\pm
}=u_{i}^{\pm }\mathrm{f}_{A}^{i}$.\newline
(\textbf{2}) integrate eq(\ref{bfa}) with respect the harmonic variables $%
u_{i}^{\pm }$. \newline
Concerning the first point, we have to solve the constraint eqs on the
auxiliary fields $\Delta _{A}^{---}$ and $B_{\mu A}^{-}$. These calculations
are technical and lengthy. Below, we give the main lines.\newline
For the case of the auxiliary field $\Delta _{A}^{---}$, the constraint eq
reads as follows 
\begin{equation}
\begin{tabular}{llll}
$\left[ \partial ^{++}-\lambda \left( \mathrm{\tilde{\phi}}^{+}T^{I}\mathrm{%
\phi }^{+}\right) T_{I}\right] \mathrm{\phi }^{+}$ & $=$ & $0$ & ,%
\end{tabular}
\label{uh}
\end{equation}%
where $\partial ^{++}$ is as in eq(\ref{dpp}) and where we have set $%
d_{IJ}T^{J}=T_{I}$. Eq(\ref{uh}) can be easily solved as%
\begin{equation}
\begin{tabular}{llll}
$\phi _{A}^{+}\left( x,u\right) $ & $=$ & $u_{i}^{+}\left( e^{\lambda \zeta }%
\mathrm{f}_{A}^{i}\left( x\right) \right) $ & ,%
\end{tabular}%
\end{equation}%
with 
\begin{equation}
\begin{tabular}{llll}
$\zeta $ & $=$ & $\sum\limits_{I=1}^{n}\zeta _{I}T^{I}$ & ,%
\end{tabular}
\label{xi}
\end{equation}%
and%
\begin{equation}
\begin{tabular}{llll}
$\zeta _{I}$ & $=$ & $u_{(i}^{+}u_{k)}^{-}\overline{\mathrm{f}}^{i}T_{I}%
\mathrm{f}^{k}$ & .%
\end{tabular}
\label{xj}
\end{equation}%
Thanks to the conservation laws which equate $\overline{\mathrm{\phi }}%
^{+}T_{I}\mathrm{\phi }^{+}=\overline{\mathrm{f}}^{+}T_{I}\mathrm{f}^{+}$
and makes eq(\ref{uh}) solvable.\newline
For the case of the auxiliary field $B_{\mu A}^{-}$, the constraint eq reads
as follows%
\begin{equation}
\begin{tabular}{llll}
$\left[ \partial ^{++}-\lambda \left( \mathrm{\tilde{\phi}}^{+}T^{I}\mathrm{%
\phi }^{+}\right) T_{I}\right] B_{\mu A}^{-}-\lambda \mathrm{K}_{\mu }%
\mathrm{\phi }_{A}^{+}$ & $=$ & $2\partial _{\mu }\mathrm{\phi }^{+}$ & ,%
\end{tabular}
\label{bm}
\end{equation}%
where we have set%
\begin{equation}
\begin{tabular}{llll}
$\mathrm{K}_{\mu }$ & $=$ & $\left( \tilde{B}_{\mu }^{-}T^{I}\mathrm{\phi }%
^{+}+\mathrm{\tilde{\phi}}^{+}T^{I}B_{\mu }^{-}\right) T_{I}$ & .%
\end{tabular}
\label{am}
\end{equation}%
Eq(\ref{am}) satisfies the useful property%
\begin{equation}
\begin{tabular}{llll}
$\partial ^{++}\mathrm{K}_{\mu }$ & $=$ & $2iT_{I}\partial _{\mu }\left( 
\mathrm{\tilde{\phi}}^{+}T^{I}\mathrm{\phi }^{+}\right) $ & .%
\end{tabular}%
\end{equation}%
To solve eq(\ref{bm}), we first set%
\begin{equation}
\begin{tabular}{llll}
$B_{\mu }^{-}$ & $=$ & $e^{\lambda \zeta }C_{\mu }^{-}$ & ,%
\end{tabular}%
\end{equation}%
with $\zeta $ as in eqs(\ref{xi}-\ref{xj}), and then look for $C_{\mu }^{-}$%
. Lengthy, but straightforward, calculations lead to,%
\begin{equation*}
\begin{tabular}{llll}
$C_{\mu }^{-A}$ & $=$ & $2\partial _{\mu }\mathrm{f}^{-A}+\lambda \mathcal{F}%
_{I}^{J}\upsilon _{\mu }^{I}Q_{J}^{-A}+\lambda Q_{J}^{+A}\tilde{Q}%
_{B}^{-J}\left( \partial _{\mu }\mathrm{f}^{-B}\right) +\lambda
Q_{J}^{+A}Q^{-BJ}\left( \partial _{\mu }\mathrm{\tilde{f}}_{B}^{-}\right) $
& $,$%
\end{tabular}%
\end{equation*}%
where we have set%
\begin{equation}
\begin{tabular}{llll}
$Q_{I}^{iaA}$ & $=$ & $\left( T_{I}\right) _{C}^{A}\mathrm{f}^{iC},\quad 
\overline{Q}_{jB}^{I}=\overline{\mathrm{f}}_{jD}\left( T^{I}\right) _{B}^{D}$
& .%
\end{tabular}%
\end{equation}%
Putting the expression of $\mathrm{\phi }^{+}$ and $B_{\mu }^{-}$ back into (%
\ref{bfa}) and integrating with respect to the harmonic variables, we end
with the following result:

\begin{equation}
\begin{tabular}{llll}
$2h_{kC}^{lD}$ & $=$ & $+2\delta _{k}^{l}\left( \delta _{C}^{D}+\frac{%
\lambda }{2}\overline{\mathrm{f}}_{iA}\mathrm{f}^{iB}d_{IJ}\left(
T^{I}\right) _{C}^{A}\left( T^{J}\right) _{B}^{D}\right) $ &  \\ 
&  & $-\lambda \mathcal{F}_{I}^{J}\left( \left[ \overline{\mathrm{f}}_{kA}%
\mathrm{f}^{lB}d_{JL}\left( T^{I}\right) _{C}^{A}\left( T^{L}\right) _{B}^{D}%
\right] \right) $ &  \\ 
&  & $-\lambda \mathcal{F}_{I}^{J}\left[ \left( \mathcal{E}_{J}^{K}\overline{%
\mathrm{f}}_{A}^{l}\mathrm{f}_{k}^{D}d_{IL}\left( T^{L}\right)
_{C}^{A}\left( T^{K}\right) _{B}^{D}\right) \right] $ & 
\end{tabular}%
,  \label{a}
\end{equation}

and

\begin{equation}
\begin{tabular}{llll}
$g^{kClD}$ & $=$ & $+\frac{\lambda }{2}\mathcal{F}_{I}^{J}\left[ \mathrm{f}%
^{kA}\mathrm{f}^{lB}\left( d_{JL}\left( T^{L}\right) _{A}^{C}\left(
T^{I}\right) _{B}^{D}\right) \right] $ &  \\ 
&  & $+\frac{\lambda }{2}\mathcal{F}_{I}^{J}\left( \mathrm{f}^{lA}\mathrm{f}%
^{kB}\left[ \mathcal{E}_{J}^{K}d_{IL}\left( T^{L}\right) _{A}^{C}\left(
T^{K}\right) _{B}^{D}\right] \right) $ &  \\ 
&  & $+\frac{\lambda }{2}\mathcal{F}_{I}^{J}\left[ \mathrm{f}^{jA}\mathrm{f}%
^{ik}\left( \mathcal{E}_{J}^{K}d_{IL}\left( T^{L}\right) _{A}^{C}\left(
T^{K}\right) _{B}^{D}\right) \varepsilon ^{kl}\varepsilon _{ij}\right] $ & 
\end{tabular}%
,  \label{b}
\end{equation}

as well as%
\begin{equation}
\begin{tabular}{llll}
$\overline{g}_{kClD}$ & $=$ & $+\frac{\lambda }{2}\mathcal{F}_{I}^{J}\left[ 
\overline{\mathrm{f}}_{lB}\overline{\mathrm{f}}_{kA}^{I}d_{JL}\left(
T^{L}\right) _{D}^{B}\left( T^{I}\right) _{C}^{A}\right] $ &  \\ 
&  & $+\frac{\lambda }{2}\mathcal{F}_{I}^{J}\left( \overline{\mathrm{f}}_{lA}%
\overline{\mathrm{f}}_{kB}\left[ \mathcal{E}_{J}^{K}d_{IL}\left(
T^{L}\right) _{C}^{A}\left( T^{K}\right) _{D}^{B}\right] \right) $ &  \\ 
&  & $+\frac{\lambda }{2}\mathcal{F}_{I}^{J}\left[ \overline{\mathrm{f}}_{jA}%
\overline{\mathrm{f}}_{iB}\left( \mathcal{E}_{J}^{K}d_{IL}\left(
T^{L}\right) _{C}^{A}\left( T^{K}\right) _{D}^{B}\right) \varepsilon
_{kl}\varepsilon ^{ij}\right] $ & .%
\end{tabular}
\label{c}
\end{equation}

\ \ \ \newline
In these relations, the space time scalars $\mathrm{f}^{iA}$ are in the
bi-fundamental of $SU\left( 2\right) \times U\left( n\right) $ and the
matrices $\mathcal{E}_{I}^{J}$ and $\mathcal{F}_{J}^{K}$ are given by 
\begin{equation}
\begin{tabular}{llll}
$\mathcal{E}_{I}^{J}$ & $=$ & $\left[ \delta _{I}^{J}+\lambda \left( 
\overline{\mathrm{f}}T_{I}T^{J}\mathrm{f}\right) \right] $ & , \\ 
$\mathcal{E}_{I}^{J}\mathcal{F}_{J}^{K}$ & $=$ & $\delta _{I}^{K}$ & .%
\end{tabular}
\label{ef}
\end{equation}%
Notice that for the leading case $n=1$, eq(\ref{ef}) reduces to%
\begin{equation}
\begin{tabular}{llll}
$\mathcal{E}$ & $=$ & $1+\lambda \overline{\mathrm{f}}\mathrm{f}$ & , \\ 
$\mathcal{F}$ & $=$ & $\frac{1}{1+\lambda \overline{\mathrm{f}}\mathrm{f}}$
& .%
\end{tabular}%
\end{equation}%
It is not difficult to check that the metric components $h_{kC}^{lD}$, $%
g^{kClD}$ and $\overline{g}_{kClD}$ reduce exactly to the metric terms $%
h_{k}^{l}$, $g^{kl}$ and $\overline{g}_{kl}$ of the real four dimensional
Taub-NUT geometry. The explicit computations and the technical details from
eq(\ref{ln}) to eq(\ref{c}) as well as other results will be reported
elsewhere.

\section{Conclusion and discussions}

In this paper, we have derived the explicit field expression of the metric $%
\hat{G}_{ab}^{IJ}=\hat{G}_{ab}^{IJ}\left( \sigma ,\phi \right) $ of the
scalar manifold $\boldsymbol{M}_{n}^{6D,N=2}$ of generic non chiral $%
\mathcal{N}=2$ supergravity in six dimensional space time. Generally, the
moduli space $\boldsymbol{M}_{n}^{6D,N=2}$ is given by 
\begin{equation}
\begin{tabular}{llll}
$SO\left( 1,1\right) \times \frac{SO\left( 4,n\right) }{SO\left( 4\right)
\times SO\left( n\right) }$ & , & $n\geq 1$ & .%
\end{tabular}%
\end{equation}%
It is a generic real $\left( \emph{1+4n}\right) $- dimensional manifold
parameterized by the local real field coordinates $\left( \sigma ,\phi
_{I}^{a}\right) $ where 
\begin{equation}
\begin{tabular}{llllllll}
$\phi _{I}^{a}$ & $\sim $ & $\sum\limits_{i,k=1}^{2}\mathcal{\sigma }%
_{ik}^{a}\phi _{I}^{ik}$ & , & $\overline{\left( \phi _{I}^{a}\right) }$ & $%
= $ & $\phi _{I}^{a}$ & ,%
\end{tabular}%
\end{equation}%
with $a=1,2,3,4$ and $I=1,...,20$. \newline
To get the explicit field expression of the metric $\hat{G}_{ab}^{IJ}$, we
have to work a little bit harder as we need various tools and several steps.
To that purpose, we have first reviewed some specific aspects on \emph{10D}
type IIA superstring on K3 and developed useful ingredients to approach $%
\hat{G}_{ab}^{IJ}$; such as the duality relation (\ref{gf}) and the
quaternionic potential (\ref{qp}). \newline
One of the basic tools that we have used to determine $\hat{G}_{ab}^{IJ}$ is
the harmonic superspace (\emph{HSS}) method; which is known to be a powerful
method for building quaternionic metrics \textrm{\cite{IS,6,ShS,8}}. In this
regards, recall that like in the case of \emph{4D} $\mathcal{N}=1$
supersymmetry and Kahler geometry, the \emph{HSS} metrics building method
relies on the link between \emph{6D} $\mathcal{N}=1$\ supersymmetry and
quaternionic geometry. \newline
Below, we summarize the main steps that we have used to derive $\hat{G}%
_{ab}^{IJ}$ or equivalently, the components $h_{iA}^{jB},$ $\overline{g}%
_{iAjB}$ and $g^{iAjB}$:\newline
(\textbf{1}) First, start from the non linear sigma model field action $%
\mathcal{S}_{b}$ associated with the scalar manifold\ $\frac{SO\left(
1,1\right) \times SO\left( 4,n\right) }{SO\left( 4\right) \times SO\left(
n\right) }$, 
\begin{equation}
\begin{tabular}{llll}
$\mathcal{S}_{b}$ & $\sim $ & $\int d^{6}x\sqrt{-\det \text{g}}g^{\mu \nu
}\left( \partial _{\mu }\sigma \partial _{\nu }\sigma -e^{-\sigma }\partial
_{\mu }\phi _{I}^{a}\partial _{\nu }\phi _{J}^{b}G_{ab}^{IJ}\right) $ & ,%
\end{tabular}
\label{sb}
\end{equation}%
where the scalar field $\sigma $ is the dilaton parameterizing the factor $%
SO\left( 1,1\right) $; and where the metric component 
\begin{equation}
\begin{tabular}{llll}
$G_{ab}^{IJ}$ & $=$ & $G_{ab}^{IJ}\left( \phi \right) $ & ,%
\end{tabular}
\label{54}
\end{equation}%
has no dependence in $\sigma $. The metric $G_{ab}^{IJ}$ concerns then the
real \emph{4n} dimensional quaternionic manifold $\frac{SO\left( 4,n\right) 
}{SO\left( 4\right) \times SO\left( n\right) }$. The minus sign in front of
the second term of the right hand side is required by the flat limit $%
G_{ab}^{IJ}\rightarrow -\delta ^{IJ}\delta _{ab}$.\newline
(\textbf{2}) Second, focus on the term $G_{ab}^{IJ}$ (\ref{54}) by freezing
the dilaton $\sigma $ in eq(\ref{sb}); that is by setting $d\sigma =0$ in $%
\mathcal{S}_{b}$. This restriction allows to use rigid supersymmetry in 
\emph{6D} to deal with the real \emph{4n} dimensional quaternionic metric $%
G_{ab}^{IJ}$. \newline
(\textbf{3}) Third, use the following duality relation, 
\begin{equation}
\begin{tabular}{llll}
$\phi _{i}^{kI}$ & $=$ & $\overline{\mathrm{f}}_{k}T^{I}\mathrm{f}^{i}$ & ,%
\end{tabular}
\label{fff}
\end{equation}%
extending the typical relation $\phi _{i}^{k}=\overline{\mathrm{f}}_{k}%
\mathrm{f}^{i}$. This remarkable duality relation maps the real 4- vector $%
\phi _{i}^{k}$ (and in general $\phi _{i}^{kI}$) to the hermitian $2\times 2$
matrix $\overline{\mathrm{f}}_{k}\mathrm{f}^{i}$ (resp. $\overline{\mathrm{f}%
}_{k}T^{I}\mathrm{f}^{i}$ ). \newline
The complex fields $\mathrm{f}^{iA}$ are in the $\left( \text{\b{2}},\text{%
\b{n}}\right) $ bi-fundamental of $SU\left( 2\right) \times U\left( n\right) 
$ and the $T^{I}$'s are the commuting Cartan generators of $U\left( n\right) 
$. \newline
The price to pay for the field change (\ref{fff}) is the symmetries of the
moduli since the scalar manifold $\frac{SO\left( 4,n\right) }{SO\left(
4\right) \times SO\left( n\right) }$ gets mapped to,%
\begin{equation}
\begin{tabular}{llll}
$\frac{SU\left( 2,n\right) }{SU\left( 2\right) \times U\left( n\right) }$ & $%
\subset $ & $\frac{SO\left( 4,n\right) }{SO\left( 4\right) \times SO\left(
n\right) }$ & .%
\end{tabular}%
\end{equation}%
The use of the field coordinates $\mathrm{f}^{iA}$ and $\overline{\mathrm{f}}%
_{iA}$ allows to split the metric $G_{ab}^{IJ}\left( \phi \right) $ into the
following form,%
\begin{equation}
G=\left( 
\begin{array}{cc}
h_{iA}^{jB} & g^{iAjB} \\ 
\overline{g}_{iAjB} & h_{iA}^{jB}%
\end{array}%
\right) .  \label{me}
\end{equation}%
Eq(\ref{fff}) permits as well to exhibit manifestly the $U^{n}\left(
1\right) $ gauge invariance of the Maxwell-matter sector of the 6D $\mathcal{%
N}=2$ supergravity. The $U^{n}\left( 1\right) $ gauge change 
\begin{equation}
\begin{tabular}{llllllll}
$\mathrm{f}^{i}$ & $\rightarrow $ & $e^{i\vartheta }\mathrm{f}^{i}$ & , & $%
\vartheta $ & $=$ & $\sum\limits_{I=1}^{20}\vartheta _{I}T^{I}$ & ,%
\end{tabular}%
\end{equation}%
leaves invariant eq(\ref{fff}); thanks to the relation $T^{I}e^{i\vartheta
}=e^{i\vartheta }T^{I}$.\newline
(\textbf{4}) Then, use the rigid \emph{6D} harmonic superspace formalism
with $d\sigma =0$; and think about eq(\ref{sb}) as the bosonic part of the
following \emph{HSS}\ superfield action,%
\begin{equation}
\begin{tabular}{lll}
$\mathcal{S}_{n}$ & $=$ & $\int d^{6}xd^{4}\theta ^{+}du\left[ \mathrm{%
\tilde{\Phi}}^{+}D^{++}\mathrm{\Phi }^{+}-\frac{\lambda }{2}\left( \mathrm{%
\tilde{\Phi}}^{+}T^{I}\mathrm{\Phi }^{+}\right) d_{IJ}\left( \mathrm{\tilde{%
\Phi}}^{+}T^{J}\mathrm{\Phi }^{+}\right) \right] $%
\end{tabular}%
,
\end{equation}%
where the interaction term 
\begin{equation}
\mathcal{L}_{int}^{+4}=\frac{\lambda }{2}\sum\limits_{I,J=1}^{20}\left( 
\mathrm{\tilde{\Phi}}^{+}T^{I}\mathrm{\Phi }^{+}\right) d_{IJ}\left( \mathrm{%
\tilde{\Phi}}^{+}T^{J}\mathrm{\Phi }^{+}\right) ,
\end{equation}%
has been derived in section 3; eq(\ref{qp}). In the above relation, the 
\emph{HSS} superfield $\mathrm{\Phi }^{+}$ is the off shell representation
of the hypermultiplet; and $\lambda $ is a real coupling constant which may
be interpreted as the $S^{4}$ area of the $AdS_{2}\times S^{4}$ near horizon
geometry of the 6D black hole.\newline
The next steps are to perform the following:\newline
(\textbf{i}) integrate the superfield action $\mathcal{S}_{n}$ with respect
to the Grassmann variables $\theta ^{+}$. This brings the action $\mathcal{S}%
_{n}$\ to the form%
\begin{equation}
\begin{tabular}{llll}
$\mathcal{S}_{n}$ & $=$ & $\int d^{6}xdu\mathcal{L}\left( \phi
^{+},B^{-},\Delta ^{---},u^{\pm }\right) $ & ,%
\end{tabular}%
\end{equation}%
(\textbf{ii}) eliminate the auxiliary fields $B^{-}$ and $\Delta ^{---}$ of
the off shell hypermultiplets $\mathrm{\Phi }^{+}$ (\ref{ex}-\ref{ey})
through their eqs of motion. This reduces $\mathcal{S}_{n}$ further to%
\begin{equation}
\begin{tabular}{llll}
$\mathcal{S}_{n}$ & $=$ & $\int d^{6}xdu\mathcal{L}\left( \mathrm{f}%
^{iA},u_{i}^{\pm }\right) $ & .%
\end{tabular}%
\end{equation}%
(\textbf{iii}) then integrate the above relation with respect to the
harmonic variables $u_{i}^{\pm }$. \newline
After doing all these steps, we end with the following component fields
action 
\begin{equation}
\begin{tabular}{lll}
$\mathcal{S}_{n}$ & $=$ & $\frac{1}{2}\int d^{6}x\left( 2h_{iA}^{jB}\partial
_{\mu }\mathrm{f}^{iA}\partial ^{\mu }\overline{\mathrm{f}}_{jB}+\overline{g}%
_{iAjB}\partial _{\mu }\mathrm{f}^{iA}\partial ^{\mu }\mathrm{f}%
^{jB}+g^{iAjB}\partial _{\mu }\overline{\mathrm{f}}_{iA}\partial ^{\mu }%
\overline{\mathrm{f}}_{jB}\right) $%
\end{tabular}%
,
\end{equation}%
from which we read the metric components 
\begin{equation}
\begin{tabular}{llll}
$h_{iA}^{jB}$ & $=$ & $h_{iA}^{jB}\left( \mathrm{f,}\overline{\mathrm{f}}%
\right) $ & , \\ 
$\overline{g}_{iAjB}$ & $=$ & $\overline{g}_{iAjB}\left( \mathrm{f,}%
\overline{\mathrm{f}}\right) $ & , \\ 
$g^{iAjB}$ & $=$ & $g^{iAjB}\left( \mathrm{f,}\overline{\mathrm{f}}\right) $
& .%
\end{tabular}
\label{bl}
\end{equation}%
These relations are explicitly given by eqs(\ref{a}-\ref{b}-\ref{c}).\newline
We end this study by making two comments: one regarding the effective
potential of the 6D black hole. The other concerns the uplifting to 7D.

\subsection{Effective potential\emph{\ }$\mathcal{V}_{BH}^{6D,N=2}$}

With this metric (\ref{bl}) at hand, we can use it to study the black hole
effective potential (\ref{ve}) which we rewrite in the form, 
\begin{equation}
\begin{tabular}{lllll}
$\mathcal{V}_{BH}^{6D,N=2}$ & $=$ & $\sum_{a,b=1}^{4}K^{ab}\left( e^{2\sigma
}\left[ Z_{a}Z_{b}+\sum\limits_{I,J=1}^{n}G_{ab}^{IJ}Z_{I}Z_{J}\right]
\right) $ & $\geq 0$ & ,%
\end{tabular}%
\end{equation}%
where we have use the factorization 
\begin{equation}
\mathcal{K}_{ab}\left( \sigma ,\phi \right) =e^{-2\sigma }K_{ab}\left( \phi
\right) .
\end{equation}%
The above relation can be also rewritten by using the complex coordinates 
\textrm{f}$^{iA}$ and $\overline{\mathrm{f}}_{iA}$; that is $\mathcal{V}%
_{BH}^{6D,N=2}=\mathcal{V}_{BH}\left( \mathrm{f}^{iA},\overline{\mathrm{f}}%
_{iA}\right) $. \newline
Below, we will mainly focus our attention on describing the way to compute
the quantities $Z_{a}$, $Z_{I}$ and $\mathcal{K}_{ab}$ involved in $\mathcal{%
V}_{BH}^{6D,N=2}$ for the case of the black hole in \emph{10D} type IIA
superstring on K3.\newline
The \emph{four} $Z_{a}$'s are the geometric central charges of the 6D $%
\mathcal{N}=2$ supersymmetry. They are the dressed charges \textrm{\cite%
{BDSS,F1,F2}} associated with the fluxes of the \emph{four} gauge field
strengths of the supergravity multiplet (\ref{gr}). A way to define the $%
Z_{a}$'s is as follows 
\begin{equation}
\begin{tabular}{llll}
$Z_{a}$ & $=$ & $\int_{K3}\mathcal{H}_{2}\wedge \boldsymbol{J}_{a}$ & .%
\end{tabular}%
\end{equation}%
where $\boldsymbol{J}_{a}\sim \boldsymbol{J}_{ij}$ is the quaternionic
2-form given by eqs(\ref{qua}-\ref{qui}) and the real 2-form $\mathcal{H}%
_{2} $ is given by, 
\begin{equation}
\begin{tabular}{llll}
$\mathcal{H}_{2}$ & $=$ & $\int_{S_{\infty }^{2}}\mathcal{F}_{4}$ & ,%
\end{tabular}%
\end{equation}%
with $\mathcal{F}_{4}$ being the the gauge invariant real 4-form field
strength $d\mathcal{C}_{3}$ of \emph{10D} type IIA superstring. The real
2-sphere $S_{\infty }^{2}$ belongs to the \emph{6D} space time.\newline
The \emph{twenty} dressed charges $Z_{I}$ are the so called matter central
charges associated with the fluxes of the \emph{twenty} gauge field
strengths of the Maxwell sector of the supergravity theory. Like for $Z_{a}$%
, the $Z_{I}$'s may be also defined as $\int_{K3}\mathcal{H}_{2}\wedge 
\boldsymbol{J}_{I}$ where the real 2-form $\boldsymbol{J}_{I}$ captures the
stringy and the geometric deformation moduli. Moreover, as in the case of
special Kahler geometry, the $Z_{I}$'s can be also given by the covariant
derivatives,%
\begin{equation}
\begin{tabular}{llll}
$Z_{I}$ & $=$ & $D_{I}^{a}Z_{a}$ & .%
\end{tabular}%
\end{equation}%
More explicit relations regarding the structure of the covariant derivatives
can be found in \textrm{\cite{BFKM,S1}}. \newline
Concerning the real symmetric matrix $\mathcal{K}_{ab}$; it is given by the
following intersections 
\begin{equation}
\begin{tabular}{llllll}
$\mathcal{K}_{ab}$ & $=$ & $\int_{K3}\boldsymbol{J}_{a}\wedge \boldsymbol{J}%
_{b}$ & , & $a,b=1,...,4$ & .%
\end{tabular}%
\end{equation}%
It factorizes as $\mathcal{K}_{ab}\left( \sigma ,\phi \right) =e^{-2\sigma
}K_{ab}\left( \phi \right) $ where $e^{-2\sigma }$ is associated with the
factor \emph{SO}$\left( 1,1\right) $ and the real symmetric field matrix $%
K_{ab}\left( \phi \right) $ describing the contribution of quaternionic
manifold $\frac{SO\left( 4,n\right) }{SO\left( 4\right) \times SO\left(
n\right) }$. \newline
The next steps are to compute the explicit expression of these quantities in
the complex coordinate frame; write down the attractor eqs of the 6D black
hole (black membrane) and then look for their solutions.

\subsection{Uplifting to 7D}

\qquad Here we develop a way to get the explicit field expression of the
metric $G_{UV}^{\left( 7D\right) }=G_{UV}\left( \sigma ,\xi \right) $, $%
U,V=1,...,58$, of the generic scalar manifolds 
\begin{equation}
\begin{tabular}{llll}
$\boldsymbol{M}_{n}^{7D,N=2}$ & $=$ & $\frac{SO\left( 3,n\right) }{SO\left(
3\right) \times SO\left( n\right) }\times SO\left( 1,1\right) $ & ,%
\end{tabular}%
\end{equation}%
of the 7D $\mathcal{N}=2$ supergravity models with $n$ Maxwell
supermultiplets. The factor $SO\left( 1,1\right) $ is parameterized by the
field isosinglet $\sigma $ and the factor $\frac{SO\left( 3,n\right) }{%
SO\left( 3\right) \times SO\left( n\right) }$ by the \emph{19} real
isotriplets field coordinates 
\begin{equation}
\xi ^{\alpha u},\text{ \ \ }\alpha =1,2,3,\text{ \ \ }u=1,...,19,
\end{equation}%
so that the metric of $\boldsymbol{M}_{n}^{7D,N=2}$ reads as 
\begin{equation}
dl^{2}=\left( d\sigma \right) ^{2}-e^{-2\sigma }\sum\limits_{\alpha ,\beta
=1}^{3}\sum\limits_{u,v=1}^{19}G_{\alpha \beta }^{uv}d\xi _{u}^{\alpha }d\xi
_{u}^{\beta }.
\end{equation}%
Notice that the $SO\left( 3\right) $ 3- vectors $\xi ^{\alpha u}$ can be
also put in the form of $SU\left( 2\right) $ isotriplets $\xi ^{\left(
ij\right) u}$ by using the homomorphism $SO\left( 3\right) \simeq SU\left(
2\right) $.\newline
To determine the explicit field expression of the metric component $%
G_{\alpha \beta }^{uv}$, we start from our result (\ref{a}-\ref{b}-\ref{c})
on the metric $G_{ab}^{IJ}$ of 
\begin{equation}
\begin{tabular}{llll}
$\boldsymbol{M}_{n+1}^{6D,N=2}$ & $=$ & $\frac{SO\left( 4,n+1\right) }{%
SO\left( 4\right) \times SO\left( n+1\right) }\times SO\left( 1,1\right) $ & 
.%
\end{tabular}%
\end{equation}%
and impose the appropriate constraint eqs that map%
\begin{equation}
\begin{tabular}{llll}
$\boldsymbol{M}_{n+1}^{6D,N=2}$ & $\rightarrow $ & $\boldsymbol{M}%
_{n}^{7D,N=2}$ & .%
\end{tabular}%
\end{equation}%
Below, we focus on the case $n=19$ corresponding to embedding \emph{7D}
(resp. \emph{6D}) supergravity in \emph{11D} M- theory (resp. \emph{10D}
type IIA superstring) on K3. \newline
First consider the local coordinates $\left( \sigma ,\phi _{I}^{ij}\right) $
of the moduli space $\boldsymbol{M}_{20}^{6D,N=2}$ used in section 3. Then
notice that the field coordinate variables $\phi _{I}^{ij}$ can be usually
decomposed into a symmetric and antisymmetric parts as follows, 
\begin{equation}
\begin{tabular}{llllll}
$\phi _{I}^{ij}$ & $=$ & $\phi _{I}^{\left[ ij\right] }+\phi _{I}^{\left(
ij\right) }$ & , & $I=1,...,19$ & , \\ 
$\phi _{I}^{ij}$ & $=$ & $\phi _{I}^{\left[ ij\right] }+\chi ^{\left(
ij\right) }$ & , & $I=20$ & ,%
\end{tabular}
\label{fii}
\end{equation}%
with%
\begin{equation}
\begin{tabular}{llllll}
$\phi _{I}^{\left[ ij\right] }$ & $=$ & $\chi _{I}^{0}\varepsilon ^{ij}$ & ,
& $I=1,...,19$ & , \\ 
$\phi _{I}^{\left[ ij\right] }$ & $=$ & $\phi ^{0}\varepsilon ^{ij}$ & , & $%
I=20$ & .%
\end{tabular}
\label{fij}
\end{equation}%
Notice also that from the view of the real dimensions of the scalar
manifolds $\boldsymbol{M}_{19}^{6D,N=2}$ and $\boldsymbol{M}_{19}^{7D,N=2}$
namely,%
\begin{equation}
\begin{tabular}{llll}
$\dim \boldsymbol{M}_{20}^{6D,N=2}$ & $=$ & $1+\left( 4\times 19+4\right) $
& , \\ 
$\dim \boldsymbol{M}_{19}^{7D,N=2}$ & $=$ & $1+3\times 19$ & ,%
\end{tabular}%
\end{equation}%
the uplifting from \emph{6D} to \emph{7D} requires imposing \emph{23}
constraint eqs on the local field variables $\phi _{I}^{ij}$ (\ref{fii}).
This number should be thought of as 
\begin{equation}
\begin{tabular}{llll}
$23$ & $=$ & $19+3+1$ & , \\ 
& $=$ & $22+1$ & ,%
\end{tabular}%
\end{equation}%
as given by eqs(\ref{19}-\ref{23}).\newline
Physically, the constraint eqs we have to impose correspond to:\newline
(\textbf{i}) switching off the \emph{22}\ moduli associated with the B-
field fluxes. The \emph{22} undesired moduli are precisely given by the 
\emph{19} isosinglets $\chi _{I}^{0}$ and the isotriplet $\chi ^{\left(
ij\right) }$ of eq(\ref{fii}).\newline
(\textbf{ii}) Identifying the volume of K3, captured by\ the isosinglet $%
\phi ^{0}$ as in eq(\ref{19}), with the field $\sigma $ parameterizing the $%
SO\left( 1,1\right) $ factor. \newline
A way to put these constraint eqs is to simply set 
\begin{equation}
\begin{tabular}{llllll}
$\chi _{I}^{0}$ & $=$ & $0$ & , & $I=1,...,19$ & , \\ 
$\chi ^{\left( ij\right) }$ & $=$ & $0$ &  &  & ,%
\end{tabular}
\label{cs1}
\end{equation}%
together with%
\begin{equation}
\begin{tabular}{llll}
$\phi ^{0}$ & $\sim $ & $e^{\sigma }$ & .%
\end{tabular}
\label{cs2}
\end{equation}%
Substituting into eq(\ref{fii}), we get%
\begin{equation*}
\begin{tabular}{llllll}
$\xi _{u}^{\left( ij\right) }$ & $=$ & $\phi _{u}^{\left( ij\right) }$ & , & 
$u=1,...,19$ & .%
\end{tabular}%
\end{equation*}%
The constraint eqs(\ref{cs1}-\ref{cs2}) can be also read and solved in terms
of the complex field coordinates $\mathrm{f}^{iA}$ and $\overline{\mathrm{f}}%
_{iA}$. We have, 
\begin{equation}
\begin{tabular}{lllllll}
$\chi _{I}^{0}$ & $=$ & $\sum\limits_{A,B=1}^{20}\overline{\mathrm{f}}%
_{iA}\left( T_{I}\right) _{B}^{A}\mathrm{f}^{iB}$ & $=0$ & , & $I=1,...,19$
& ,%
\end{tabular}
\label{rea}
\end{equation}%
and%
\begin{equation}
\begin{tabular}{llllll}
$\phi ^{0}$ & $=$ & $\lambda \sum\limits_{A=1}^{20}\overline{\mathrm{f}}_{iA}%
\mathrm{f}^{iA}$ & $\sim $ & $e^{\sigma }$ & .%
\end{tabular}%
\end{equation}%
Eqs(\ref{rea}) can be solved by specifying the $U^{n}\left( 1\right) $
matrix generators $T_{I}$ of the $U\left( n\right) $ symmetry. Taking the $%
T_{I}$'s as 
\begin{equation}
\begin{tabular}{llllll}
$T_{I}$ & $=$ & $\varrho _{I}-\varrho _{I+1}$ & , & $I=1,...,19$ & , \\ 
$Tr\left( T_{I}\right) $ & $=$ & $0$ & , & $I=1,...,19$ & ,%
\end{tabular}%
\end{equation}%
and%
\begin{equation}
\begin{tabular}{llll}
$T_{20}$ & $\simeq $ & $\varrho _{1}+...+\varrho _{20}$ & ,%
\end{tabular}%
\end{equation}%
with $\varrho _{I}$ being the matrix projectors ($\varrho _{I}^{2}=\varrho
_{I}$) on the vector basis $\left\{ e_{I}\right\} $ of the underlying 20-
dimensional space ($\varrho _{K}e_{I}=\delta _{KI}e_{I}$), it follows from
the constraint eqs $\chi _{I}^{0}=0$ and the realization (\ref{rea}) that%
\begin{equation}
\begin{tabular}{llllllll}
$\overline{\mathrm{f}}_{i1}\mathrm{f}^{i1}$ & $=$ & $\overline{\mathrm{f}}%
_{i2}\mathrm{f}^{i2}$ & $=$ & $...$ & $=$ & $\overline{\mathrm{f}}_{i20}%
\mathrm{f}^{i20}$ & .%
\end{tabular}
\label{fr}
\end{equation}%
These relations tell us that the uplifting to 7D requires that all complex
fields $\mathrm{f}^{iA}$ should have the same norm.\newline
Putting this solution back into the second relation of (\ref{rea}), we
obtain for each complex isodoublet $\mathrm{f}^{iA}$ the following norm,%
\begin{equation}
\begin{tabular}{llll}
$\lambda \overline{\mathrm{f}}_{iA}\mathrm{f}^{iA}$ & $=$ & $\frac{1}{20}%
e^{\sigma }$ & .%
\end{tabular}
\label{fs}
\end{equation}%
To conclude, the metric $G_{\alpha \beta }^{uv}$ of the moduli space $\frac{%
SO\left( 3,19\right) }{SO\left( 3\right) \times SO\left( 19\right) }\times
SO\left( 1,1\right) $ of the \emph{7D} $\mathcal{N}=2$ supergravity is
recovered from eqs(\ref{bl}) by imposing the constraint relations (\ref{fr}-%
\ref{fs}). This result applies as well for the generic scalar manifolds $%
\frac{SO\left( 3,n\right) }{SO\left( 3\right) \times SO\left( n\right) }%
\times SO\left( 1,1\right) $ with $n\geq 1$.

\begin{acknowledgement}
:\qquad\ \ \newline
This research work is supported by the programme PROTARS\ D12/25/CNRST.
\end{acknowledgement}

\section{Appendix: Generalities on \emph{HSS} method}

Harmonic superspace method has been first introduced for solving the problem
of a manifestly off shell superspace formulation of \emph{4D} $\mathcal{N}=2$
extended supersymmetric Yang Mills $\mathrm{\cite{1}}$ and \emph{4D} $%
\mathcal{N}=2$ supergravity theories; for a review see $\mathrm{\cite{2}}$,
see also \textrm{\cite{3,30}}. \newline
This method has been extended as well to other dimensions and for different
purposes, in particular to \emph{2D} supersymmetric field theories with
eight supercharges $\mathrm{\cite{4}}$. It has been used also for
approaching different matters; in particular for studying \emph{4D}
Euclidean Yang-Mills and gravitational instantons $\mathrm{\cite{5}}$ and
refs therein, in the HK metrics building \textrm{\cite{IS,6,ShS,8}}, in
dealing with the analysis of singularities of so called HyperKahler
Calabi-Yau manifolds used in type II superstring compactifications $\mathrm{%
\cite{9}}$ and in topological string on conifold \textrm{\cite{30}}.

In the \emph{HSS} formulation of \emph{4D} $\mathcal{N}=2$ extended (or
equivalently \emph{6D} $\mathcal{N}=1$) supersymmetric theory, the ordinary
superspace with $SU\left( 2\right) $ R-symmetry, 
\begin{equation}
z^{M}=\left( x^{\mu },\theta _{a}^{i},\overline{\theta }_{\dot{a}%
}^{i}\right) ,\qquad i=1,2\quad a,\text{ }\dot{a}=1,2,
\end{equation}%
gets mapped into the harmonic superspace%
\begin{equation}
z^{M}=\left( Y^{m},\theta _{a}^{-},\overline{\theta }_{\dot{a}%
}^{-},u_{i}^{\pm }\right) ,
\end{equation}%
with analytic supercoordinates $Y^{m}$ as%
\begin{equation}
Y^{m}=\left( y^{\mu },\theta _{a}^{+},\overline{\theta }_{\dot{a}%
}^{+}\right) ,
\end{equation}%
and 
\begin{eqnarray}
y^{\mu } &=&x^{\mu }+i\left( \theta ^{+}\sigma ^{\mu }\overline{\theta }%
^{-}+\theta ^{-}\sigma ^{\mu }\overline{\theta }^{+}\right) ,  \notag \\
\theta _{i}^{+} &=&u_{i}^{+}\theta _{a}^{i},\qquad \overline{\theta }_{\dot{a%
}}^{+}=u_{i}^{+}\overline{\theta }_{\dot{a}}^{i},
\end{eqnarray}%
where $u_{i}^{\pm }$ are the harmonic variables satisfying, amongst others,
the relation $u_{i}^{-}u^{+i}=1$. \newline
In \emph{HSS}, the \emph{4D} $\mathcal{N}=2$ (or equivalently the \emph{6D} $%
\mathcal{N}=1$) hypermultiplets are represented by the analytic harmonic
superfunction 
\begin{equation}
\Phi ^{+}=\Phi ^{+}\left( Y,u\right) .
\end{equation}%
This \emph{HSS} function carry one positive Cartan-Weyl charge%
\begin{equation}
\left[ D^{0},\Phi ^{+}\right] =\Phi ^{+},
\end{equation}%
where the U$\left( 1\right) $ charge operator D$^{0}$ is given by%
\begin{equation}
D^{0}=\partial ^{0}+\left( \theta ^{+}\frac{\partial }{\partial \theta ^{+}}+%
\overline{\theta }^{+}\frac{\partial }{\partial \overline{\theta }^{+}}%
\right) -\left( \theta ^{-}\frac{\partial }{\partial \theta ^{-}}+\overline{%
\theta }^{-}\frac{\partial }{\partial \overline{\theta }^{-}}\right) .
\end{equation}%
The superfield $\Phi ^{+}$ satisfies as well the following analyticity
condition, 
\begin{equation}
D_{a}^{+}\Phi ^{+}=0,\qquad \overline{D}_{\dot{a}}^{+}\Phi ^{+}=0,\qquad a,%
\text{ }\dot{a}=1,2,
\end{equation}%
where $D_{a}^{+}=u_{i}^{\pm }D_{a}^{i}$ and $\overline{D}_{\dot{a}%
}^{+}=u_{i}^{\pm }\overline{D}_{\dot{a}}^{i}$ are the supersymmetric
covariant derivatives of the \emph{4D} $\mathcal{N}=2$ superalgebra. They
can be thought of as%
\begin{equation}
D_{a}^{+}=\frac{\partial }{\partial \theta ^{-a}},\qquad \overline{D}_{\dot{a%
}}^{+}=\frac{\partial }{\partial \overline{\theta }^{-\dot{a}}}.
\end{equation}%
The $\theta ^{+}$- expansion of $\Phi ^{+}$ reads as 
\begin{equation}
\Phi ^{+}\left( Y,u\right) =\phi ^{+}+\theta ^{+2}F^{-}+\overline{\theta }%
^{+2}G^{-}+i\theta ^{+a}\overline{\theta }^{+\dot{a}}B_{a\dot{a}}^{-}+\theta
^{+2}\overline{\theta }^{+2}\Delta ^{---},  \label{sp}
\end{equation}%
where we have dropped out fermions for simplicity. \newline
Notice that the components, 
\begin{equation}
F^{-}=F^{-}\left( x,u\right) ,\text{ \quad }G^{-}=G^{-}\left( x,u\right) ,%
\text{ \quad }B_{a\dot{a}}^{-}=B_{a\dot{a}}^{-}\left( x,u\right) ,
\label{ax}
\end{equation}%
are auxiliary fields scaling as a mass squared; i.e \emph{mass}$^{2}$. In
the 6D \emph{HSS}\ formulation, these auxiliary fields combine altogether
into a complex 6D\ vector $B_{\left[ \hat{\alpha}\hat{\beta}\right] }^{-}$.
The remaining one namely, 
\begin{equation}
\Delta ^{---}=\Delta ^{---}\left( x,u\right) ,  \label{ay}
\end{equation}%
is also an auxiliary field; but scaling as \emph{mass}$^{3}$. All these
fields are needed to have off shell supersymmetric theory. We also have 
\begin{equation}
\tilde{\Phi}^{+}\left( Y,u\right) =\tilde{\phi}^{+}+\theta ^{+2}\tilde{G}%
^{-}+\overline{\theta }^{+2}\tilde{F}^{-}+i\theta ^{+a}\overline{\theta }^{+%
\dot{a}}\tilde{B}_{a\dot{a}}^{-}+\theta ^{+2}\overline{\theta }^{+2}\tilde{%
\Delta}^{---}  \label{ps}
\end{equation}%
where $\left( \sim \right) $ stands for the twild conjugation preserving\
the harmonic analyticity \cite{1}. $\left( \sim \right) =\left( \overline{%
\ast }\right) $ is a combination of the complex conjugation (bar) and the $%
\left( \ast \right) $- conjugation of the $U\left( 1\right) $ subsymmetry of 
$SU\left( 2\right) $ R-symmetry.\ Moreover, the component fields 
\begin{equation}
\mathrm{F}^{q}=\mathrm{F}^{q}\left( y,u\right) ,
\end{equation}%
with Cartan charge $q$, can be also expanded in a harmonic series as follows:%
\begin{equation}
\mathrm{F}^{q}\left( y,u\right) =\sum_{n=0}^{\infty
}u_{(i_{1}}^{+}u_{i_{2}}^{+}...u_{i_{n+q}}^{+}u_{j_{1}}^{-}...u_{j_{n})}^{-}%
\mathrm{F}^{\left( i_{1}...i_{n+q}j_{1}...j_{n}\right) }\left( y\right) ,
\end{equation}%
where we have taken $q\geq 0$. Notice in passing that one of the key
difficulties in \emph{HSS} method is to determine the right dependence of
the component field $\mathrm{F}^{q}\left( y,u\right) $ in terms of the
dynamical scalars $\mathrm{f}^{i}\left( y\right) $ and $\overline{\mathrm{f}}%
_{i}\left( y\right) $ of the hypermultiplets. \newline
A quite similar formulation is valid for the gauge supermultiplet $V_{4D}^{%
\mathcal{N}=2}$ (or equivalently \emph{6D} $\mathcal{N}=1$ vector
supermultiplet $V_{6D}^{+N=1}$). It is described by the analytic
prepotential 
\begin{equation}
\begin{tabular}{llll}
$V^{++}$ & $=$ & $V^{++}\left( Y\mathbf{,}u\right) $ & , \\ 
$D_{a}^{+}V^{++}$ & $=$ & $0$ & , \\ 
$\overline{D}_{\dot{a}}^{+}V^{++}$ & $=$ & $0$ & ,%
\end{tabular}%
\end{equation}%
carrying two positive Cartan-Weyl charge,%
\begin{equation}
\left[ D^{0},V^{++}\right] =2V^{++}.
\end{equation}%
The $\theta ^{+}$- expansion of $V^{++}$ reads, in the Wess-Zumino gauge as
follows 
\begin{equation}
V^{++}\left( Y,u\right) =\theta ^{+2}C+\overline{\theta }^{+2}\overline{C}%
+\theta ^{+a}\overline{\theta }^{+\dot{a}}A_{a\dot{a}}+\theta ^{+2}\overline{%
\theta }^{+2}P^{--}.
\end{equation}%
Notice that in the 6D \emph{HSS}, the field components $C$, $\overline{C}$
and $A_{a\dot{a}}$ combine altogether to form a 6D gauge field $A_{\left[ 
\hat{\alpha}\hat{\beta}\right] }$.\newline
Besides describing the gauge dynamics, the gauge superfield $V^{++}$ is also
used to covariantize the \emph{HSS} derivative, 
\begin{equation}
D^{++}=\partial ^{++}-2i\theta ^{+}\sigma ^{\mu }\overline{\theta }%
^{+}\partial _{\mu }-2i\theta ^{+^{2}}\frac{\partial }{\partial \overline{%
\tau }}-2i\overline{\theta }^{+2}\frac{\partial }{\partial \tau },
\label{hsd}
\end{equation}%
where we have set $\tau =x^{4}+ix^{5}$. The gauge covariant harmonic
derivative reads as 
\begin{equation}
\mathcal{D}^{++}=D^{++}-iV^{++},\qquad \tilde{V}^{++}=V^{++}.
\end{equation}%
Under HSS gauge transformation of hypermultiplet superfields%
\begin{equation}
\Phi ^{+\prime }=e^{i\Lambda }\Phi ^{+},
\end{equation}%
where $\Lambda $ is a real analytic gauge superparameter, U$\left( 1\right) $
gauge covariance%
\begin{equation}
\mathcal{D}^{++}\Phi ^{+\prime }=e^{i\Lambda }\mathcal{D}^{++}\Phi ^{+}
\end{equation}%
requires that 
\begin{equation}
V^{++\prime }=V^{++}+D^{++}\Lambda .
\end{equation}%
Notice that setting $\theta ^{+}=0$, the harmonic derivative $D^{++}$ and
its partners 
\begin{equation}
\begin{tabular}{llll}
$D^{--}$ & $=$ & $\partial ^{--}-2i\theta ^{-}\sigma ^{\mu }\overline{\theta 
}^{-}\partial _{\mu }-2i\theta ^{-^{2}}\frac{\partial }{\partial \overline{%
\tau }}-2i\overline{\theta }^{-2}\frac{\partial }{\partial \tau }$ & $,$ \\ 
$D^{0}$ & $=$ & $\left[ D^{++},D^{-}\right] $ & ,%
\end{tabular}%
\end{equation}%
coincide exactly with the usual $\partial ^{++}$, $\partial ^{--}$ and $%
\partial ^{0}$ generators and obey the same su$\left( 2\right) $ algebra (%
\ref{al}) namely, 
\begin{eqnarray}
\left[ D^{0},D^{++}\right] &=&2D^{++},\quad  \notag \\
\left[ D^{0},D^{--}\right] &=&-2D^{--},\quad \\
\left[ D^{++},D^{--}\right] &=&D^{0}.  \notag
\end{eqnarray}%
Notice finally that $D^{++}$, $D^{--}$ and $D^{0}$ are real under the
conjugation $\sim $, i.e $\tilde{D}^{++}=D^{++}$, $\tilde{D}^{0}=D^{0}$ and $%
\tilde{D}^{--}=D^{--}$ and they play a crucial role in the \emph{HSS}
formulation of \emph{4D} $\mathcal{N}=2$ (6D $\mathcal{N}=1$) supersymmetric
field theory.

\section{Appendix B: Geometrical approach}

In this appendix, we review briefly the main lines of an alternative
geometric method for dealing with the field dynamics in non chiral $\mathcal{%
N}=2$ supergravity in \emph{6D},\emph{\ }including the scalar field
couplings considered in this paper. This geometrical construction, which has
been developed in a series of papers \textrm{\cite{10}}-\textrm{\cite{R},}
is a powerful method based on generalized Maurer-Cartan equations and the
solving of the superspace Bianchi identities of the so called $\widehat{%
\mathrm{F}}_{4}$ supergravity containing the \emph{6D}$\ \mathcal{N}=2$
superalgebra as a subsymmetry. Though beyond the objective of our paper, we
think that the geometric approach deserves a comment as it is an alternative
way to the harmonic superspace method that we have developed in the present
study. \newline
Before going into technical details, it is instructif to start by describing
the main result. According to the studies \textrm{\cite{10}}-\textrm{\cite%
{12}, }the geometric method leads to\ the following supersymmetric gauge
invariant component field action,%
\begin{equation}
\begin{tabular}{ll}
$\mathcal{S}_{D=6}^{N=2}=\int d^{6}x\sqrt{-g}\left[ \mathcal{L}_{kin}+%
\mathcal{L}_{{\small cs}}\right] +\int d^{6}x\sqrt{-g}\mathcal{L}_{{\small p}%
}+$ {\small higher spinor terms} & .%
\end{tabular}
\label{act}
\end{equation}%
The kinetic term $\mathcal{L}_{kin}$ and the Chern-Simons one $\mathcal{L}_{%
{\small cs}}$ are given by,%
\begin{equation}
\begin{tabular}{lll}
$\mathcal{L}_{kin}+\mathcal{L}_{CS}=$ & $-\frac{1}{4}R_{6}-\frac{1}{8}%
e^{2\sigma }\mathcal{N}_{\Lambda \Sigma }F_{\mu \nu }^{\Lambda }F^{\Sigma
\mu \nu }+\frac{3}{64}e^{4\sigma }H_{\mu \nu \rho }H^{\mu \nu \rho }$ &  \\ 
& $+\frac{i}{2}\overline{\psi }_{\mu i}\gamma ^{\mu \nu \rho }D_{\nu }\psi
_{\rho }^{i}-2i\overline{\chi }_{i}\gamma ^{\mu }D_{\mu }\chi ^{i}+\frac{i}{8%
}\overline{\lambda }_{i}^{I}\gamma ^{\nu }D_{\nu }\lambda _{I}^{i}$ &  \\ 
& $\partial _{\mu }\sigma \partial ^{\mu }\sigma -\frac{1}{4}%
\sum\limits_{y,z=1}^{4n}\left(
P_{y}^{I0}P_{I0z}+\sum\limits_{r=1}^{3}P_{y}^{Ir}P_{Irz}\right) \partial
_{\mu }\phi ^{y}\partial ^{\mu }\phi ^{z}$ &  \\ 
& $-\frac{1}{64}\epsilon ^{\mu \nu \rho \sigma \lambda \tau }B_{\mu \nu
}F_{\rho \sigma }^{\Lambda }F_{\lambda \tau }^{\Sigma }\eta _{\Lambda \Sigma
}$ & .%
\end{tabular}
\label{kin}
\end{equation}%
The Pauli term $\mathcal{L}_{{\small p}}$ as well as the higher fermionic
terms are lengthy and highly untrivial; their explicit expression can be
found in \textrm{\cite{10}}. In the Lagrangian density (\ref{kin}), one
recognizes the pure scalar fields term 
\begin{equation*}
\partial _{\mu }\sigma \partial ^{\mu }\sigma
-\sum\limits_{y,z=1}^{4n}\left(
P_{y}^{I0}P_{I0z}+\sum\limits_{r=1}^{3}P_{y}^{Ir}P_{Irz}\right) \partial
_{\mu }\phi ^{y}\partial ^{\mu }\phi ^{z},
\end{equation*}%
with $P^{I0}=\sum_{y=1}^{4n}P_{y}^{I0}d\phi ^{y}$ being the vielbein of the
coset $\frac{SO\left( 4,n\right) }{SO\left( 4\right) \times SO\left(
n\right) }$, and should be then compared with eq(\ref{1}). The gauge field
coupling $\mathcal{N}_{\Lambda \Sigma }$ reads, in terms of the field matrix 
$L_{\Lambda \Sigma }$ parameterizing the duality group $SO\left( 4,n\right) $%
, as follow: 
\begin{equation}
\begin{tabular}{ll}
$\mathcal{N}_{\Lambda \Sigma }=\sum\limits_{a=1}^{4}L_{a\Lambda }L_{\Sigma
}^{a}-\sum\limits_{I=1}^{n}L_{I\Lambda }L_{\Sigma }^{I}$ & .%
\end{tabular}%
\end{equation}%
The other fields appearing in (\ref{kin}) are the usual gauge invariant
components of the non chiral $\mathcal{N}=2$ supergravity fields; some of
them were already discussed in previous sections of our paper, others will
be defined below. \newline
Notice that the geometrical approach we are describing here has been first
considered in \textrm{\cite{R}} for the case of pure supergravity. It has
been further developed in \textrm{\cite{10,11,120} }by implementing the
matter couplings and gaugings. The geometrical approach has been also used
for the study of special features of \emph{AdS/CFT} correspondence; in
particular the points regarding the two following issues: \textbf{(i)} the
relation between the $SU(2)$ gauging coupling constant $g$ and the inverse $%
AdS_{6}$ radius $4m$; see eq(\ref{gm}). \textbf{(ii)} the description of the
Higgs phenomenon by which the gravitational two-form $B_{\mu \nu }$ becomes
massive; see eq(\ref{ab}). Below, we give a general sketch of this
construction by following the analysis given in the above mentioned works.

\textbf{(1) Pure supergravity }\newline
Using the same convention notations as in \textrm{\cite{10,11}}, the field
content of the pure \emph{6D} non chiral $N=2$ supergravity multiplet in a 
\emph{Poincar\'{e} background} reads as, 
\begin{equation*}
\begin{tabular}{llll}
$V_{\mu }^{a},$ $B_{\mu \nu },$ $A_{\mu }^{\alpha },$ $e^{\sigma }$ & $;$ & $%
\psi _{\mu }^{A},$ $\psi _{\mu }^{\dot{A}},$ $\chi ^{A},$ $\chi ^{\dot{A}}$
& ,%
\end{tabular}%
\end{equation*}%
with $V_{\mu }^{a}$ being the \emph{6D} space time vielbein, with indices $%
a,b=0,...,5$ being the Lorentz flat indices, $\mu ,\nu =0,...,5$ the
corresponding world indices and:\newline
the fields\textbf{\ }$\psi _{\mu }^{A},\,\ \psi _{\mu }^{\dot{A}}$ are
respectively the left-handed and the right-handed four- component gravitino
fields with $A,$ $\dot{A}=1,2$ transforming under the two factors of the $R$%
- symmetry group $O(4)\simeq SU(2)_{L}\times SU(2)_{R}$. \newline
$B_{\mu \nu }$ is the antisymmetric field with field strength $\mathcal{H}%
_{\mu \nu \rho }$. \newline
$A_{\mu }^{\alpha }$ ($\alpha =0,1,2,3$), are the four real \emph{6D} vector
gauge fields with field strength $\mathcal{F}_{\mu \nu }^{\alpha }$. \newline
$\chi ^{A},\chi ^{\dot{A}}$ describing respectively the left-handed and the
right-handed spin $\frac{1}{2}$ four components dilatinos, and finally $%
e^{\sigma }$ denotes the dilaton.

\emph{Unified description of Poincar\'{e} and AdS backgrounds}\newline
With AdS$_{6}$/CFT$_{5}$ correspondence in mind, it is interesting to start
by describing some useful tools. The point is that the description of the
spinors of the gravity multiplet in terms of left-handed and right-handed
projection holds only in a \emph{Poincar\'{e} background}. There, the $%
SO(1,5)$ Weyl spinors are 4-dimensional (4-dim) with $R$-symmetry group as $%
SU(2)_{L}\times SU(2)_{R}$. In the $AdS_{6}$ background, the chiral
projection cannot however be defined; one is then restricted to use rather
8-dim\emph{\ }pseudo-Majorana spinors. The $SO(2,5)$ spinors are 8-dim
pseudo real and the above $R$-symmetry gets reduced to the $SU(2)$ diagonal
subgroup of $SU(2)_{L}\times SU(2)_{R}$. \newline
To study in a \emph{unique setting} both Poincar\'{e} and $AdS_{6}$ vacua,
it is then convenient to use from the 8-dim pseudo-Majorana spinors even in
a Poincar\'{e} framework. The pseudo-Majorana condition on the gravitino
1-forms $\psi _{A}=\psi _{\mu A}dx^{\mu }$ is as follows $\left( \psi
_{A}\right) ^{\dagger }\gamma ^{0}=\overline{(\psi _{A})}=\epsilon ^{AB}\psi
_{B}^{T}$. The indices $A,B=1,2$ of the spinor fields $\psi _{A},\,\chi _{A}$
transform in the fundamental of the diagonal subgroup $SU(2)$ contained in $%
SU(2)_{L}\times SU(2)_{R}$. We also have the following properties for the
8-dim antisymmetric gamma matrices: $\gamma ^{7}=i\gamma ^{0}\gamma
^{1}\gamma ^{2}\gamma ^{3}\gamma ^{4}\gamma ^{5}$, $\gamma _{7}^{T}=-\gamma
_{7}$ and $(\gamma _{7})^{2}=-1$.

$\widehat{\mathrm{F}}_{4}$ \emph{superalgebra }\newline
Using the fact that the $\widehat{F}_{4}$ \emph{supergroup} \textrm{\cite{13}%
-\cite{15} }has as bosonic subsymmetry the $SO(2,5)\times SU(2)$ group
product and simply following the studies of refs \textrm{\cite{10,11,120}},
we can construct $\widehat{F}_{4}$ supergravity by using supergeometry
techniques. To that purpose, we start by considering the 1-form gauge fields,%
\begin{equation*}
\begin{tabular}{llll}
$\left\{ 
\begin{array}{c}
V^{a}=V_{\mu }^{a}dx^{\mu } \\ 
\omega ^{ab}=\omega _{\mu }^{ab}dx^{\mu }%
\end{array}%
\right. $ & and & $A^{r}=A_{\mu }^{r}dx^{\mu }$, $r=1,2,3$ & ,%
\end{tabular}%
\end{equation*}%
respectively associated to the $AdS_{6}$ and $SU(2)$ algebras. The 1- forms $%
V^{a}$ and $\omega ^{ab}$ are dual to the $so(2,5)$ generators $\left(
P_{a},M_{ab}\right) $ and $A^{r}$ to the $su(2)$ generators $T_{r}$. The $%
\left( P_{a},M_{ab}\right) $ matrices satisfy the following $AdS_{6}$
commutation relations,%
\begin{equation}
\begin{tabular}{llll}
$\lbrack P_{a},P_{b}]$ & $=$ & $8m^{2}M_{ab}$ & , \\ 
$\lbrack M_{ab},P_{c}]$ & $=$ & $\frac{1}{2}\left( \eta _{ac}P_{b}-\eta
_{bc}P_{a}\right) $ & , \\ 
$\lbrack M_{ab},M_{cd}]$ & $=$ & $\frac{1}{2}\left( \eta _{bc}M_{ad}+\eta
_{ad}M_{bc}-\eta _{bd}M_{ac}-\eta _{ac}M_{bd}\right) $ & ,%
\end{tabular}
\label{fof}
\end{equation}%
while the $T_{r}$'s obey: 
\begin{equation*}
\begin{tabular}{llll}
$\lbrack T^{s},T^{t}]$ & $=$ & $ig\varepsilon ^{str}T_{r}$ & , \\ 
$\lbrack T^{s},P_{a}]$ & $=$ & $[T^{s},M_{ab}]=0$ & .%
\end{tabular}%
\end{equation*}%
In above relations, the number $g$ is the gauge coupling constant of $SU(2)$
and $m$ is related to the $AdS_{6}$ radius of $\frac{SO(2,5)}{SO(1,5)}$ by 
\begin{equation*}
m=(2R_{AdS})^{-1}.
\end{equation*}%
To construct the full $\widehat{\mathrm{F}}_{4}$ \emph{superalgebra}, the
minimal extension of the conformal group $SO\left( 2,5\right) $ or
equivalently of the $AdS_{{\small 6}}$ group, one introduces the
pseudo-Majorana spinor charges $Q_{A\alpha }=\left( Q_{A1},...,Q_{A8}\right) 
$ and enlarges the 
\begin{equation*}
so(2,5)\oplus su(2)
\end{equation*}%
algebra to the full $\widehat{\mathrm{F}}_{4}$ \emph{superalgebra}. The
simplest way to get $\widehat{F}_{4}$ is to proceed as follows: \textbf{(i)}
Start from the Maurer-Cartan equation of $SO(2,5)\times SU(2)$ gauge
symmetry,%
\begin{equation}
\begin{tabular}{llll}
$\mathcal{D}V^{a}\equiv dV^{a}-\omega ^{ab}V_{b}$ & $=$ & $0$ & , \\ 
$\mathcal{R}^{ab}+4m^{2}V^{a}V^{b}$ & $=$ & $0$ & , \\ 
$dA^{r}+\frac{g}{2}\varepsilon ^{rst}A_{s}A_{t}$ & $=$ & $0$ & ,%
\end{tabular}
\label{ma}
\end{equation}%
with $\mathcal{R}^{ab}\equiv d\omega ^{ab}-\omega ^{ac}\wedge \omega
_{c}^{b} $. \textbf{(ii)} Enlarge these eqs by implementing the spinor
1-forms $\psi ^{A\alpha }=\psi _{\mu }^{A\alpha }dx^{\mu }$ dual to the
fermionic generators $Q_{A\alpha }$. \newline
Following \textrm{\cite{10,11,12}}, the minimal extension of (\ref{ma}) is
given by:%
\begin{equation}
\begin{tabular}{llll}
$\mathcal{D}V^{a}$ & $=$ & $\frac{i}{2}\overline{\psi }_{A}\gamma _{a}\psi
^{A}$ & , \\ 
$\mathcal{R}^{ab}+4m^{2}V^{a}V^{b}+$ & $=$ & $-m\overline{\psi }_{A}\gamma
_{ab}\psi ^{A}$ & , \\ 
$dA^{r}+\frac{g}{2}\varepsilon ^{rst}A_{s}A_{t}$ & $=$ & $i\overline{\psi }%
_{A}\psi _{B}\sigma ^{rAB}$ & , \\ 
$D\psi _{A}$ & $=$ & $im\gamma _{a}\psi _{A}V^{a}$ & ,%
\end{tabular}
\label{mu}
\end{equation}%
where $D$ is the $SO(1,5)\times SU(2)$ covariant derivative which acts on
spinors as follows,%
\begin{equation*}
\begin{tabular}{llll}
$D\psi _{A}$ & $=$ & $\mathcal{D}\psi _{A}-\frac{i}{2}\sigma
_{AB}^{r}A_{r}\psi ^{B}$ & $,$ \\ 
& $=$ & $d\psi _{A}-\frac{1}{4}\gamma _{ab}\omega ^{ab}\psi _{A}-\frac{i}{2}%
\sigma _{AB}^{r}A_{r}\psi ^{B}$ & ,%
\end{tabular}%
\end{equation*}%
with $\sigma ^{r\left( AB\right) }=\frac{1}{2}\epsilon ^{BC}\sigma _{C}^{rA}$
with $\sigma _{B}^{rA}$\ denoting the usual Pauli matrices. One of the
remarkable properties of eqs (\ref{mu}) is that their closure under \emph{%
d-differentiation} is ensured only if the following relation holds 
\begin{equation}
g=3m.  \label{gm}
\end{equation}%
The graded commutation relation of the $\widehat{F}_{4}$ supergroup are
obtained by using the standard identity $d\omega (X,Y)=\frac{1}{2}\left(
X(\omega (Y))-Y(\omega (X))-\omega \lbrack X,Y]\right) $ and the duality
relations%
\begin{equation*}
\begin{tabular}{llll}
$\omega ^{ab}(M_{cd})$ & $=$ & $\delta _{cd}^{ab}$ & , \\ 
$V^{a}(P_{b})$ & $=$ & $\delta _{b}^{a}$ & , \\ 
$\psi ^{A\alpha }(Q_{B\beta })$ & $=$ & $\delta _{b}^{a}\delta _{\beta
}^{\alpha }$ & ,%
\end{tabular}%
\end{equation*}%
The super- commutation relations read, in addition to eqs(\ref{fof}), as
follows 
\begin{equation*}
\begin{tabular}{llll}
$\left\{ \overline{Q}_{A\alpha },Q_{B\beta }\right\} $ & $=$ & $-i\epsilon
_{AB}\left( \gamma ^{a}\right) _{\alpha \beta }P_{a}+4i\delta _{\alpha \beta
}T_{\left( AB\right) }+m\epsilon _{AB}\left( \gamma ^{ab}\right) _{\alpha
\beta }M_{ab}$ & , \\ 
$\lbrack M_{ab},\overline{Q}_{A\beta }]$ & $=$ & $-\frac{1}{4}\overline{Q}%
_{A\alpha }\left( \gamma _{ab}\right) _{\alpha \beta }$ & , \\ 
$\lbrack P_{a},\overline{Q}_{A\beta }]$ & $=$ & $im\overline{Q}_{A\alpha
}\left( \gamma _{a}\right) _{\alpha \beta }$ & , \\ 
$\lbrack T_{\left( AB\right) },\overline{Q}_{C\alpha }]$ & $=$ & $\frac{i}{2}%
g\left( \overline{Q}_{A\alpha }\delta _{BC}+\overline{Q}_{B\alpha }\delta
_{AC}\right) $ & ,%
\end{tabular}%
\end{equation*}%
where $T_{\left( AB\right) }$ stands for $T_{r}\sigma _{AB}^{r}$ and
obviously $g=3m$ as an outcome of the graded Jacobi identities.

\emph{Free Differential Algebra} (FDA.)\newline
Using results obtained in \textrm{\cite{11,12,FDA}, }the supersymmetric
Maurer-Cartan eqs(\ref{mu}) keep the same form when we pass from the $%
\widehat{F}_{4}$ supergroup to the superspace coset $\frac{\widehat{F}_{4}}{%
SO(1,5)\times SU(2)}$, which contains $AdS_{6}$ as bosonic submanifold. The
previous 1-forms%
\begin{equation*}
\begin{tabular}{llllllll}
$V^{a}$ & , & $\omega ^{ab}$ & , & $\psi _{A}$ & , & $A^{r}$ & ,%
\end{tabular}%
\end{equation*}%
become now \emph{superfield 1-forms} describing the vacuum configuration in
superspace whose bosonic subspace is $AdS_{6}$. Notice that on the ordinary
space-time, recovered by setting odd superspace dimension $\theta =0$ in the
superfields 1-forms, the background vacuum fields have precisely as $dx^{\mu
}$ components the following expressions:%
\begin{equation*}
\begin{tabular}{llllll}
$V_{\mu }^{a}=\delta _{\mu }^{a}$ & , & $\psi _{A\mu }=0$ & , & $\left(
\omega _{\mu }^{ab},A_{\mu }^{r}\right) =${\small pure\thinspace gauge} & .%
\end{tabular}%
\end{equation*}%
Notice also that because of the absence of the 2-form $B$ and of the 1-form $%
A^{0}$ superfields, eqs(\ref{mu}) cannot describe the supersymmetric vacuum
of the full $\widehat{F}_{4}$ supergravity theory. To overcome this
difficulty, we use the \emph{Free Differential Algebra} (FDA.) \textrm{\cite%
{FDA}} obtained from the $\widehat{F}_{4}$ Maurer-Cartan eqs (\ref{mu}) by
adding two more equations as given below,%
\begin{equation}
\begin{tabular}{llll}
$\mathcal{D}V^{a}-\frac{i}{2}\overline{\psi }_{A}\gamma _{a}\psi ^{A}$ & $=$
& $0$ & , \\ 
$\mathcal{R}^{ab}+4m^{2}V^{a}V^{b}+m\overline{\psi }_{A}\gamma _{ab}\psi
^{A} $ & $=$ & $0$ & , \\ 
$dA^{r}+\frac{g}{2}\varepsilon ^{rst}A_{s}A_{t}-i\overline{\psi }_{A}\psi
_{B}\sigma ^{rAB}$ & $=$ & $0$ & , \\ 
$D\psi _{A}-im\gamma _{a}\psi _{A}V^{a}$ & $=$ & $0$ & ,%
\end{tabular}
\label{d}
\end{equation}%
and%
\begin{equation}
\begin{tabular}{llll}
$dA-mB-i\overline{\psi }_{A}\gamma _{7}\psi ^{A}$ & $=$ & $0$ & , \\ 
$dB+2\overline{\psi }_{A}\gamma _{7}\gamma _{a}\psi ^{A}V^{a}$ & $=$ & $0$ & 
.%
\end{tabular}
\label{dd}
\end{equation}%
One of the very remarkable features of the FDA eqs, describing the full
supersymmetric vacuum configuration, is the appearance of the combination 
\begin{equation}
dA^{0}-mB=\frac{1}{2}dx^{\mu }\wedge dx^{\nu }\left( \partial _{\lbrack \mu
}A_{\nu ]}^{0}-mB_{\left[ \mu \nu \right] }\right) .  \label{ab}
\end{equation}%
At the dynamical level, this relation implies a Higgs phenomenon where the
2-form $B$ \emph{eats} the 1-form $A^{0}$ and acquires a non vanishing mass $%
m$.\newline
Setting $m=g=0$, one reduces the $\widehat{\mathrm{F}}_{4}$ superalgebra to
the \emph{6D}$\ \mathcal{N}=2$ superalgebra existing only in a super- Poincar%
\'{e} background. In this case, the four gauge fields 
\begin{equation*}
A^{\alpha }\equiv (A^{0},A^{r})
\end{equation*}%
transforms in the fundamental of the $R$-symmetry group $SO(4)$ \ and the
pseudo-Majorana spinors $\psi _{A},\chi _{A}$ decomposes in two chiral
spinors in such a way that all the resulting FDA is invariant under $SO(4)$.%
\newline
Moreover it is no difficult to see that no FDA exists for the cases 
\begin{equation*}
\begin{tabular}{ll}
$\left\{ 
\begin{array}{c}
m=0,\text{ \ }g\neq 0 \\ 
m\neq 0,\text{ \ }g=0%
\end{array}%
\right. $ & ,%
\end{tabular}%
\end{equation*}%
since the corresponding FDA eqs do not close anymore under $d$-
differentiation. For a supersymmetric vacuum to exist, the gauging of $SU(2)$%
, 
\begin{equation*}
g\neq 0,
\end{equation*}%
must be necessarily accompanied by the presence of the parameter $m$ which
makes the closure of the supersymmetric algebra consistent as far as the
condition (\ref{gm}) holds. Now we turn to discuss matter couplings.

\textbf{(2) Implementing matter vector multiplets}\newline
The vector multiplets of the 6D non chiral supergravity are given by the
multiplets $\left( A_{\mu }^{I},\lambda _{A}^{I},\phi ^{\alpha I}\right) $
where $\alpha =0,...,3$ and $I=1,...,n$ labeling an arbitrary number $n$ of $%
\left( A_{\mu },\lambda _{A},\phi ^{\alpha }\right) $. The $4n$ scalars $%
\phi ^{\alpha I}$ $\left( \equiv \phi ^{y}\right) $ with $y=1,...,4n$,
together with $e^{\sigma }$ of the pure supergravity multiplet, parameterize
the coset manifold 
\begin{equation*}
\begin{tabular}{lll}
$\frac{SO(4,n)}{SO(4)\times SO(n)}\times SO(1,1),$ & $G=SO(4,n)$ & .%
\end{tabular}%
\end{equation*}%
To perform the matter coupling we use the geometrical procedure whose main
lines are as follows: First, introduce the coset representative $L_{\Sigma
}^{\Lambda }$ of the matter coset manifold with $\Lambda ,\Sigma =0,\dots
,3+n$; that is a $\left( n+4\right) \times \left( n+4\right) $ matrix $%
L_{\left( n+4\right) \times \left( n+4\right) }=\left( L_{\Lambda \Sigma
}\right) $. Then, decompose the $SO(4,n)$ index $\Sigma $ with respect to $%
H=SO(4)\times SO(n)$ and put $L_{\Sigma }^{\Lambda }$ like: 
\begin{equation}
L_{\Sigma }^{\Lambda }=(L_{\alpha }^{\Lambda },L_{I}^{\Lambda })
\end{equation}%
To gauge the $SU(2)$ diagonal subgroup of $SO(4)$ as in pure supergravity,
it is useful to further decompose $L_{\alpha }^{\Lambda }$ as%
\begin{equation*}
\begin{tabular}{llll}
$L_{\alpha }^{\Lambda }=(L_{0}^{\Lambda },L_{r}^{\Lambda })$ & , & $r=1,2,3$
& .%
\end{tabular}%
\end{equation*}%
The $\left( 4+n\right) $ gravitational and matter vectors transform in the
fundamental of $SO(4,n)$, the corresponding superspace curvatures are then
labeled by the index $\Lambda $ and the covariant derivatives acting on the
spinor fields will now contain the composite connections of the $SO(4,n)$
duality group. Moreover, the $SO(4,n)$ left- invariant 1-form $\Omega
_{\Sigma }^{\Lambda }=\left( L^{-1}\right) _{\Upsilon }^{\Lambda }dL_{\Sigma
}^{\Upsilon }$ satisfying the usual Maurer-Cartan equation, 
\begin{equation*}
d\Omega _{\Sigma }^{\Lambda }+\Omega _{\Upsilon }^{\Lambda }\wedge \Omega
_{\Sigma }^{\Upsilon }=0,
\end{equation*}%
can be decomposed as follows%
\begin{equation*}
\begin{tabular}{llllll}
$R_{s}^{r}$ & $=-P_{I}^{r}\wedge P_{s}^{I}$ & , & $R_{0}^{r}$ & $%
=-P_{I}^{r}\wedge P_{0}^{I}$ & , \\ 
$R_{J}^{I}$ & $=-P_{r}^{I}\wedge P_{J}^{r}-P_{0}^{I}\wedge P_{J}^{0}$ & , & $%
\nabla P_{r}^{I}$ & $=\nabla P_{0}^{I}=0$ & ,%
\end{tabular}%
\end{equation*}%
where we have set $P_{\alpha }^{I}=\left( P_{0}^{I},P_{r}^{I}\right) \equiv
\left( \Omega _{0}^{I},\Omega _{r}^{I}\right) $ and where%
\begin{equation*}
\begin{tabular}{lll}
$R^{rs}$ & $=d\Omega _{s}^{r}+\Omega _{t}^{r}\wedge \Omega _{s}^{t}+\Omega
_{0}^{r}\wedge \Omega _{s}^{0}$ & , \\ 
$R^{r0}$ & $=d\Omega _{0}^{r}+\Omega _{t}^{r}\wedge \Omega _{0}^{t}$ & , \\ 
$R^{IJ}$ & $=d\Omega _{J}^{I}+\Omega _{K}^{I}\wedge \Omega _{J}^{K}$ & .%
\end{tabular}%
\end{equation*}%
Notice that $\left( P_{0}^{I},P_{r}^{I}\right) $ are the 1- form vielbeins
of the coset, $(\Omega ^{rs},\,\ \Omega ^{r0})$ are the connections of $%
SO(4) $ decomposed with respect to the diagonal subgroup $SU(2)\subset SO(4)$
and $(R^{rs},\,\ R^{ro})$ are the corresponding curvatures. The superspace
curvatures of the matter coupled theory read as follows%
\begin{equation}
\begin{tabular}{llllll}
$T^{A}$ & $=\mathcal{D}V^{a}-\frac{i}{2}\overline{\psi }_{A}\gamma _{a}\psi
^{A}V^{a}=0$ & , & $R^{ab}$ & $=\mathcal{R}^{ab}$ & , \\ 
$H$ & $=dB+2e^{-2\sigma }\overline{\psi }_{A}\gamma _{7}\gamma _{a}\psi
^{A}V^{a}$ & , & $R(\sigma )$ & $=d\sigma $ & , \\ 
$F^{\Lambda }$ & $=\mathcal{F}^{\Lambda }-ie^{\sigma }\overline{\psi }%
_{A}\left( L_{0}^{\Lambda }\epsilon ^{AB}\gamma _{7}+^{\sigma
}L_{r}^{\Lambda }\sigma ^{rAB}\right) \psi _{B}$ & , & $R_{0}^{I}(\phi )$ & $%
=P_{0y}^{I}d\phi ^{y}$ & , \\ 
$\rho _{A}$ & $=\mathcal{D}\psi _{A}+\frac{i}{2}\sigma _{rAB}(\frac{1}{2}%
\epsilon ^{rst}\Omega _{st}+i\gamma _{7}\Omega _{r0})\psi ^{B}$ & , & $%
R_{r}^{I}(\phi )$ & $=P_{ry}^{I}d\phi ^{y}$ & ,%
\end{tabular}
\label{cu}
\end{equation}%
together with 
\begin{equation}
\begin{tabular}{lll}
$D\chi _{A}$ & $=\mathcal{D}\chi _{A}+\frac{i}{2}\sigma _{rAB}(\frac{1}{2}%
\epsilon ^{rst}\Omega _{st}+i\gamma _{7}\Omega _{r0})\chi ^{B}$ & , \\ 
$\nabla \lambda _{IA}$ & $=\mathcal{D}\lambda _{IA}+\frac{i}{2}\sigma _{rAB}(%
\frac{1}{2}\epsilon ^{rst}\Omega _{st}+i\gamma _{7}\Omega _{r0})\lambda
_{I}^{B}+\Omega _{I}^{J}\lambda _{JA}$ & ,%
\end{tabular}
\label{cur}
\end{equation}%
These relations extend the usual curvatures 
\begin{equation}
\begin{tabular}{llllll}
$T^{a}$ & $=\mathcal{D}V^{a}-\frac{i}{2}\overline{\psi }_{A}\gamma _{a}\psi
^{A}V^{a}=0$ & , & $R^{ab}$ & $=\mathcal{R}^{ab}$ & , \\ 
$H$ & $=dB+2e^{-2\sigma }\overline{\psi }_{A}\gamma _{7}\gamma _{a}\psi
^{A}V^{a}$ & , & $\rho _{A}$ & $=D\psi _{A}$ & , \\ 
$F$ & $=dA-ie^{\sigma }\overline{\psi }_{A}\gamma _{7}\psi ^{A}$ & , & $%
R(\chi _{A})$ & $\equiv D\chi _{A}$ & , \\ 
$F^{r}$ & $=dA^{r}-ie^{\sigma }\overline{\psi }_{A}\psi _{B}\sigma ^{rAB}$ & 
, & $R(\sigma )$ & $\equiv d\sigma $ & ,%
\end{tabular}%
\end{equation}%
associated with the pure supergravity case \textrm{\cite{R}}. Notice in
passing that in eq(\ref{cu}) there appear, in the vector field strengths $%
F^{\Lambda }$, the G/H coset representatives $L_{\Sigma }^{\Lambda }$, which
intertwine between the $R$-symmetry indices $A,B$ of the gravitinos and the
indices $\Lambda ,\Sigma $ of the $\left( 4+n\right) $- dimensional $G$
representation. \newline
The Bianchi identities read as $\mathcal{D}R^{ab}=d^{2}\sigma =DP_{AB}^{I}=0$
together with 
\begin{equation*}
\begin{tabular}{lll}
$R^{ab}V_{b}-i\overline{\psi }_{A}\gamma ^{a}\rho _{B}\epsilon ^{AB}$ & $=0$
& , \\ 
$D^{2}\lambda _{A}^{I}+\frac{1}{4}R^{ab}\gamma _{ab}\lambda _{A}^{I}-\frac{i%
}{2}\sigma _{AB}^{r}(\frac{1}{2}\epsilon ^{rst}R^{st}+i\gamma
_{7}R_{r0})\lambda ^{IB}-R_{J}^{I}\lambda _{A}^{J}$ & $=0$ & , \\ 
$dH+4e^{-2\sigma }d\sigma \,\ \overline{\psi }_{A}\gamma _{7}\gamma _{a}\psi
_{B}\epsilon ^{AB}V^{a}+4e^{-2\sigma }\overline{\psi }_{A}\gamma _{7}\gamma
_{a}\rho _{B}\epsilon ^{AB}V^{a}$ & $=0$ & , \\ 
$D\rho _{A}+\frac{1}{4}R^{ab}\gamma _{ab}\psi _{A}-\frac{i}{2}\sigma
_{AB}^{r}(\frac{1}{2}\epsilon ^{rst}R^{st}+i\gamma _{7}R_{r0})\psi ^{B}$ & $%
=0$ & , \\ 
$D^{2}\chi _{A}+\frac{1}{4}R^{ab}\gamma _{ab}\chi _{A}-\frac{i}{2}\sigma
_{AB}^{r}(\frac{1}{2}\epsilon ^{rst}R^{st}+i\gamma _{7}R_{r0})\chi ^{B}$ & $%
=0$ & ,%
\end{tabular}%
\end{equation*}%
as well as%
\begin{equation*}
\begin{tabular}{lll}
$DF^{\Lambda }+id\sigma e^{\sigma }\overline{\psi }_{A}\gamma _{7}\psi
_{B}L_{[AB]}^{\Lambda }+id\sigma e^{\sigma }\overline{\psi }_{A}\psi
_{B}L_{(AB)}^{\Lambda }$ &  &  \\ 
$-2ie^{\sigma }\overline{\psi }_{A}\gamma _{7}\rho _{B}L_{[AB]}^{\Lambda
}-2ie^{\sigma }\overline{\psi }_{A}\rho _{B}L_{(AB)}^{\Lambda }+ie^{\sigma
}L_{\ \ I}^{\Lambda }\overline{\psi }^{A}\psi ^{B}P_{(AB)}^{I}$ &  &  \\ 
$+ie^{\sigma }L_{\ \ I}^{\Lambda }\overline{\psi }^{A}\gamma _{7}\psi
^{B}P_{[AB]}^{I}$ & $=0$ & ,%
\end{tabular}%
\end{equation*}%
where $P_{AB}^{I}=P_{0}^{I}\epsilon _{AB}+P_{r}^{I}\sigma _{AB}^{r}$.

\emph{Supersymmetric transformations}\newline
The solution of these Bianchi identities is highly non trivial, especially
the one regarding gravitino 1-forms where one needs cubic fermionic terms of
the form $\psi \psi \chi $; explicit results may be found in \textrm{\cite%
{10,11,12}}. Below, we content ourself to quote the solution in terms of the
supersymmetric transformations of the physical fields which, as is known,
can be written down once the parameterizations of the supercurvatures in
superspace are identified. The result is%
\begin{equation*}
\begin{tabular}{llll}
$\delta V_{\mu }^{a}$ & $=$ & $-i\overline{\psi }_{A\mu }\gamma
^{a}\varepsilon ^{A}$ & , \\ 
$\delta B_{\mu \nu }$ & $=$ & $4ie^{-2\sigma }\overline{\chi }_{A}\gamma
_{7}\gamma _{\mu \nu }\varepsilon ^{A}-4e^{-2\sigma }\overline{\varepsilon }%
_{A}\gamma _{7}\gamma _{\lbrack \mu }\psi _{\nu ]}^{A}$ & , \\ 
$\delta A_{\mu }^{\Lambda }$ & $=$ & $2e^{\sigma }\overline{\varepsilon }%
^{A}\left( L_{0}^{\Lambda }\epsilon _{AB}\gamma _{7}+L^{\Lambda r}\sigma
_{rAB}\right) \gamma _{\mu }\chi ^{B}-e^{\sigma }L_{\Lambda I}\overline{%
\varepsilon }^{A}\gamma _{\mu }\lambda ^{IB}\epsilon _{AB}$ &  \\ 
&  & $+2ie^{\sigma }\overline{\varepsilon }_{A}\left( \epsilon
^{AB}L_{0}^{\Lambda }\gamma ^{7}+L^{\Lambda r}\sigma _{r}^{AB}\right) \psi
_{B}$ & ,%
\end{tabular}%
\end{equation*}%
and 
\begin{equation*}
\begin{tabular}{llll}
$\delta \psi _{A\mu }$ & $=$ & $\mathcal{D}_{\mu }\varepsilon _{A}+\frac{1}{%
16}e^{-\sigma }[T_{[AB]\nu \lambda }\gamma _{7}-T_{(AB)\nu \lambda }](\gamma
_{\mu }^{\,\ \nu \lambda }-6\delta _{\mu }^{\nu }\gamma ^{\lambda
})\varepsilon ^{B}+$ &  \\ 
&  & $+\frac{i}{32}e^{2\sigma }H_{\nu \lambda \rho }\gamma _{7}(\gamma _{\mu
}^{\,\ \nu \lambda \rho }-3\delta _{\mu }^{\nu }\gamma ^{\lambda \rho
})\varepsilon _{A}+\frac{1}{2}\varepsilon _{A}\overline{\chi }^{C}\psi
_{C\mu }+$ &  \\ 
&  & $+\frac{1}{2}\gamma _{7}\varepsilon _{A}\overline{\chi }^{C}\gamma
^{7}\psi _{C\mu }-\gamma _{\nu }\varepsilon _{A}\overline{\chi }^{C}\gamma
^{\nu }\psi _{C\mu }+\gamma _{7}\gamma _{\nu }\varepsilon _{A}\overline{\chi 
}^{C}\gamma ^{7}\gamma ^{\nu }\psi _{C\mu }+$ &  \\ 
&  & $-\frac{1}{4}\gamma _{\nu \lambda }\varepsilon _{A}\overline{\chi }%
^{C}\gamma ^{\nu \lambda }\psi _{C\mu }-\frac{1}{4}\gamma _{7}\gamma _{\nu
\lambda }\varepsilon _{A}\overline{\chi }^{C}\gamma ^{7}\gamma ^{\nu \lambda
}\psi _{C\mu }$ & ,%
\end{tabular}%
\end{equation*}%
as well as 
\begin{equation*}
\begin{tabular}{llll}
$\delta \chi _{A}$ & $=$ & $\frac{i}{2}\gamma ^{\mu }\partial _{\mu }\sigma
\varepsilon _{A}+\frac{ie^{-\sigma }}{16}\left( T_{[AB]\mu \nu }\gamma
_{7}+T_{(AB)\mu \nu }\right) \gamma ^{\mu \nu }\varepsilon ^{B}+\frac{%
e^{2\sigma }}{32}H_{\mu \nu \lambda }\gamma _{7}\gamma ^{\mu \nu \lambda
}\varepsilon _{A}$ & , \\ 
$\delta \sigma $ & $=$ & $\overline{\chi }_{A}\varepsilon ^{A}$ & , \\ 
$\delta \lambda ^{IA}$ & $=$ & $i\left( P_{0y}^{I}\epsilon ^{AB}\gamma
^{7}-iP_{ry}^{I}\sigma ^{rAB}\right) \gamma ^{\mu }\varepsilon _{B}\partial
_{\mu }\phi ^{y}+\frac{i}{2}e^{-\sigma }T_{\mu \nu }^{I}\gamma ^{\mu \nu
}\varepsilon ^{A}$ & , \\ 
$P_{0y}^{I}\delta \phi ^{y}$ & $=$ & $\frac{1}{2}\overline{\lambda }%
_{A}^{I}\gamma _{7}\varepsilon ^{A}$ & , \\ 
$P_{ry}^{I}\delta \phi ^{y}$ & $=$ & $\frac{1}{2}\overline{\lambda }%
_{A}^{I}\varepsilon _{B}\sigma _{r}^{AB}$ & ,%
\end{tabular}%
\end{equation*}%
where 
\begin{equation}
\begin{tabular}{lll}
$T_{[AB]\mu \nu }$ & $=\epsilon _{AB}L_{0\Lambda }^{-1}F_{\mu \nu }^{\Lambda
}$ & , \\ 
$T_{(AB)\mu \nu }$ & $=\sigma _{AB}^{r}L_{r\Lambda }^{-1}F_{\mu \nu
}^{\Lambda }$ & , \\ 
$T_{I\mu \nu }$ & $=L_{I\Lambda }^{-1}F_{\mu \nu }^{\Lambda }$ & ,%
\end{tabular}%
\end{equation}%
stands for \emph{dressed} vector field strengths. \newline
Notice that in the transformation of the fermions, the cubic fermionic terms
type $(\chi \chi \varepsilon )$, $(\lambda \lambda \varepsilon )$, $(\lambda
\chi \varepsilon )$ have been omitted. An important property of the solution
presented above is that no supersymmetric $AdS_{6}$ background exists. In
the Poincar\'{e} vacuum, where all the field strengths are zero and the
scalar $\sigma $ takes an arbitrary constant value, one has,%
\begin{equation*}
\begin{tabular}{lllll}
$\delta \psi _{A\mu }=\mathcal{D}_{\mu }\varepsilon _{A}$ & , & $\delta \chi
_{A}=0$ & , & $\delta \lambda ^{IA}=0.$%
\end{tabular}%
\end{equation*}%
The solutions of the Bianchi identities give as well the equations of motion
of the physical fields which allow in turn to reconstruct the space-time
Lagrangian. The obtained component field Lagrangian is precisely the one
given by eq(\ref{act}).

\end{document}